\documentclass[final,5p,times,english,twocolumn]{elsarticle}

\usepackage{lineno}
\modulolinenumbers[5]

\usepackage{chngpage}
\usepackage{amsmath,amssymb,amsthm}
\usepackage{algorithm,algorithmic,amssymb,amstext,enumerate,amsfonts,url,multirow,array,stfloats,flushend,multirow}

\usepackage[T1]{fontenc}
\newcommand{\bm}[1]{\mbox{\boldmath{$#1$}}}

\usepackage{graphicx}
\usepackage{longtable}
\usepackage{array}
\usepackage{multirow}
\usepackage{graphicx}
\usepackage{bigstrut,bigdelim,multirow}

\usepackage{caption}
\usepackage{booktabs}
\usepackage{threeparttable}

\usepackage[colorlinks,linkcolor=blue,citecolor=blue,urlcolor=blue]{hyperref}

\usepackage{graphicx}

\usepackage{array}

\def\x{{\mathbf x}}
\def\w{{\mathbf w}}
\def\y{{\mathbf y}}
\def\u{{\mathbf u}}
\def\z{{\mathbf z}}

\def\c{{\mathbf c}}
\def\p{{\mathbf p}}

\def\h{{\mathbf h}}
\def\c{{\mathbf c}}
\def\n{{\mathbf n}}

\def\K{{\mathbf K}}

\def\I{{\mathbf I}}

\newlength\savedwidth

\newlength\savewidth

\usepackage{amsmath,color}
\usepackage{algorithm}
\usepackage{algorithmic}

\usepackage{array}

\usepackage{bigstrut,bigdelim,multirow}

\biboptions{authoryear}

\begin{document}

\begin{frontmatter}

\title{Deep ad-hoc beamforming}

\author{Xiao-Lei Zhang$^{1,2}$}
\address{$^{1}$Research \& Development Institute of Northwestern Polytechnical University in Shenzhen, Shenzhen, China\\
$^{2}$Center for Intelligent Acoustics and Immersive Communications, School of Marine Science and Technology, Northwestern Polytechnical University, China\\
 }
 \ead{xiaolei.zhang@nwpu.edu.cn}


\begin{abstract}
Far-field speech processing is an important and challenging problem. In this paper, we propose \textit{deep ad-hoc beamforming}, a deep-learning-based multichannel speech enhancement framework based on ad-hoc microphone arrays, to address the problem. It contains three novel components. First, it combines \textit{ad-hoc microphone arrays} with deep-learning-based multichannel speech enhancement, which reduces the probability of the occurrence of far-field acoustic environments significantly. Second, it groups the microphones around the speech source to a local microphone array by a supervised channel selection framework based on deep neural networks. Third, it develops a simple time synchronization framework to synchronize the channels that have different time delay. Besides the above novelties and advantages, the proposed model is also trained in a single-channel fashion, so that it can easily employ new development of speech processing techniques. Its test stage is also flexible in incorporating any number of microphones without retraining or modifying the framework. We have developed many implementations of the proposed framework and conducted an extensive experiment in scenarios where the locations of the speech sources are far-field, random, and blind to the microphones. Results on speech enhancement tasks show that our method outperforms its counterpart that works with linear microphone arrays by a considerable margin in both diffuse noise reverberant environments and point source noise reverberant environments. We have also tested the framework with different handcrafted features. Results show that although designing good features lead to high performance, they do not affect the conclusion on the effectiveness of the proposed framework.
\end{abstract}


\end{frontmatter}


 \setlength{\arraycolsep}{0.0em}

\section{Introduction}\label{sec:introduction}



Deep learning based speech enhancement has demonstrated its strong denoising ability in adverse acoustic environments  \citep{wang2018supervised}, which has attracted much attention since its first appearance \citep{wang2013towards}. Although many positive results have been observed, existing deep-learning-based speech enhancement and its applications were studied mostly with a single microphone or a conventional microphone array, such as a linear array in a portable equipment. Its performance drops when the distance between the speech source and the microphone (array) is enlarged. Finally, how to maintain the enhanced speech at the same high quality throughout an interested physical space becomes a new problem.

 \begin{figure}[t]
\centering
\includegraphics[width=6cm]{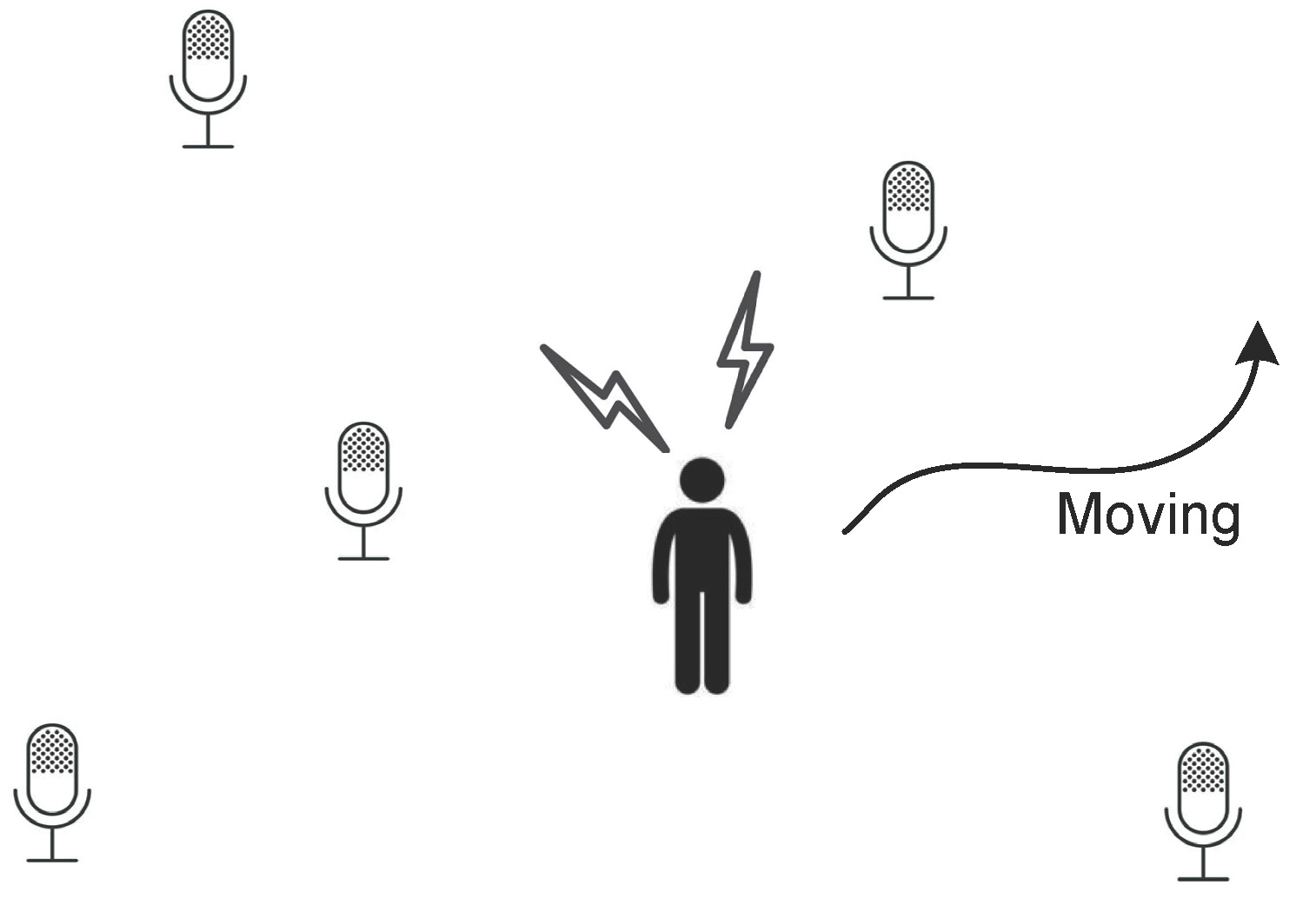}
\caption{Illustration of an ad-hoc microphone array.} \label{fig:overall}
\end{figure}

Ad-hoc microphone arrays provide a potential solution to the above problem. As illustrated in Fig. \ref{fig:overall}, an ad-hoc microphone array is a set of randomly distributed microphones. The microphones collaborate with each other. Compared to the conventional microphone arrays, an ad-hoc microphone array has the following two advantages. First, it has a chance to enhance a speaker's voice with equally good quality in a range where the array covers. Second, its performance is not limited to the physical size of application devices, e.g. cell-phones, gooseneck microphones, or smart speaker boxes. Ad-hoc microphone arrays also have a chance to be widespread in real-world environments, such as meeting rooms, smart homes, and smart cities. The research on ad-hoc microphone arrays is an emerging direction \citep{markovich2012distributed,heusdens2012distributed,zeng2014distributed,wang2015self,o2016distributed,o2014diffusion,tavakoli2016framework,jayaprakasam2017distributed,tavakoli2017distributed,zhang2018microphone,wang2018pseudo,koutrouvelis2018low}.
It contains at least the following three fundamental problems:
 \begin{itemize}
   \item \textbf{Channel selection.} Because the microphones may distribute in a large area, taking all microphones into consideration may not be the best way, since that the microphones that are far away from the speech source may be too noisy. Channel selection aims to group a handful microphones around a speaker into a local microphone array from a large number of randomly distributed microphones.

   \item \textbf{Device synchronization.} Because the microphones are distributed in different positions and maybe also different devices, their output signals may have different time delay, clock rates, or adaptive gain controllers. Device synchronization aims to synchronize the signals from the microphones, so as to fascinate its following specific applications.

   \item \textbf{Task-driven multichannel signal processing.} It aims to maximize the performance of a specific task by, e.g., adapting a multichannel signal processing algorithm designed for a conventional array to an ad-hoc array. The tasks include speech enhancement, multi-talker speech separation, speech recognition, speaker recognition, etc.
 \end{itemize}
However, current research on ad-hoc microphone arrays is still at the beginning. For example, some work discussed a so called message passing problem between microphones where it assumes that the ground-truth noise spectrum is known \citep{heusdens2012distributed} or the steering vectors from speech sources to microphones are known \citep{zeng2014distributed}. Some work focused on the channel selection problem in an ideal scenario where perfect noise estimation and voice activity detection are available \citep{zhang2018microphone}. Although some work tried to jointly conduct noise estimation and channel selection by advanced mathematical formulations, it has to make many assumptions such as the ground-truth distances between microphones, ground-truth geometry of the array, and free-space signal transmission model without reverberation \citep{o2016distributed}.

A possible explanation for the above difficulty is that an ad-hoc microphone array lacks so much important prior knowledge while contains so many interferences that we finally have little information about the array except the received signals in the extreme case. To overcome the difficulty, we may consider the way of learning the priors, parameters, or hidden variables of the array instead of making unrealistic assumptions. Supervised deep learning, which learns prior knowledge and parameters by neural networks, provides us this opportunity as it did for the supervised speech separation with the conventional microphone arrays \citep{wang2018supervised}.


 In this paper, we propose a framework named \textit{deep ad-hoc beamforming} (DAB) which brings deep learning to ad-hoc microphone arrays. It has the following four novelties:
  \begin{itemize}
  \item \textbf{A supervised channel selection framework is proposed.} It first predicts the quality of the received speech signal of each channel by a deep neural network. Then, it groups the microphones that have high speech quality and strong cross-channel signal correlation into a local microphone array. Several channel selection algorithms have been developed, including a one-best channel selection method and several \textit{N}-best channel selection methods with the positive integer $N\geq 1$ either predefined or automatically determined according to different channel selection criteria.


   \item \textbf{A simple supervised time synchronization framework is proposed.} 
       It first picks the output of the best channel as a reference signal, then estimates the relative time delay of the other channels by a traditional time delay estimator, and finally synchronize the channels according to the estimation result, where the best channel is selected in a supervised manner.

    \item \textbf{A speech enhancement algorithm is implemented as an example.} The algorithm applies the channel selection and time synchronization frameworks to deep beamforming. It is designed to demonstrate the overall effectiveness and flexibility of DAB. Its implementation is straightforward without a large modification of existing deep beamforming algorithms.

    \item \textbf{The effects of acoustic features on performance are studied.} It is known that the performance of deep-learning-based speech enhancement relies heavily on acoustic features. In this paper, we further emphasize its importance on DAB by carrying out the first study on the effects of different handcrafted features on the performance. In this study, a variant of short term Fourier transform (STFT) and the common multi-resolution cochleagram (MRCG) features are used for comparison.
  \end{itemize}
 We have conducted an extensive experimental comparison between DAB and its deep-learning-based multichannel speech enhancement counterpart with linear microphone arrays, in scenarios where the speech sources and microphone arrays were placed randomly in typical physical spaces with random time delay, and the noise sources were either diffuse noise or point source noise. Experimental results with noise-independent training show that DAB outperforms its counterpart by a large margin. Its experimental conclusion is consistent across different hyperparameter settings and handcrafted features, though a good handcrafted feature do improve the overall performance.

 The core idea of the proposed method has been released at \citep{zhang2018deep}. The main difference between the method here and \citep{zhang2018deep} is that we have added the time synchronization module, more channel selection algorithms and experiments on the point source noise environment here, which are incremental extensions that do not affect the fundamental claim of the novelty of the paper, compared to some related methods published after \citep{zhang2018deep}.


 This paper is organized as follows. Section \ref{sec:notations} presents mathematical notations of this paper. Section \ref{sec:signal_model}
 presents the signal model of ad-hoc microphone arrays. Section \ref{sec:db_ama} presents the framework of the proposed DAB. Section \ref{subsec:cs} presents the channel selection module of DAB. Section \ref{sec:problem} presents the application of DAB to speech enhancement. Section \ref{sec:no_delay} evaluates the effectiveness of the proposed method. Finally, Section \ref{sec:conclusion} concludes our findings. 

\subsection{Related work}

 Current deep-learning-based techniques employ either a single microphone or a conventional microphone array to pick up speech signals. Here, the conventional microphone array means that the microphone array is fixed in a single device. Deep-learning-based single-channel speech enhancement, e.g. \citep{wang2013towards,zhang2013denoising,lu2013speech,wang2014training,huang2015joint,xu2015regression,weninger2015speech,williamson2016complex,zhang2016deep}, employs a deep neural network (DNN), which is a multilayer perceptron with more than one nonlinear hidden layer, to learn a nonlinear mapping function from noisy speech to clean speech or its ideal time-frequency masks. This field progresses rapidly. We list some of the recent progress as follows. Phase spectrum, which was originally believed to be helpless to speech enhancement, has shown to be helpful in the deep learning methodology \citep{zheng2018phase,tan2019learning}.
 Some end-to-end speech enhancement methods have been proposed, including the representative gated residual networks \citep{tan2018gated}, fully-convolutional time-domain audio separation network \citep{luo2019conv}, and full convolutional neural networks \citep{pandey2019new}. The long-term difficulty of the nonlinear distortion of the enhanced speech for speech recognition has been overcome as well \citep{wang2019bridging}. Some solid theoretical analysis on the generalization ability of the deep learning based speech enhancement has been made \citep{qi2019theory}.

Deep-learning-based multichannel speech enhancement has two major forms. The first form \citep{jiang2014binaural} uses a microphone array as a feature extractor to extract spatial features as the input of the DNN-based single-channel enhancement.
The second form  \citep{heymann2016neural,erdogan2016improved}, which we denote bravely as \textit{deep beamforming}, estimates a monaural time-frequency (T-F) mask \citep{wang2014training,heymann2016neural,higuchi2016robust} using a single-channel DNN so that the spatial covariance matrices of speech and noise can be derived for adaptive beamforming, e.g. minimum variance distortionless response (MVDR) or generalized eigenvalue beamforming. It is fundamentally a linear method, whose output does not suffer from nonlinear distortions. Due to its success on speech recognition, it has been extensively studied, including the aspects of the integration with the spatial-clustering-based masking \citep{nakatani2017integrating}, acoustic features \citep{wang2018all}, model training \citep{xiao2017time,tu2017lstm,higuchi2018frame,zhou2018robust}, mask estimations \citep{erdogan2016improved}, post-processing \citep{zhang2017speech}, rank-1 estimation of steering vectors \citep{taherian2019deep}, etc.

The effectiveness of deep learning based speech enhancement lies strongly on acoustic features. The earliest studies take the concatenation of multiple acoustic features, such as STFT and Mel frequency captral coefficient (MFCC), as the input \citep{wang2013towards,zhang2013deep} for the sake of mining the complementary information between the features. Later on, \cite{chen2014feature} found that cochleagram feature based on gammatone filterbanks is a strong noise-robust acoustic feature, after a wide comparison between 17 acoustic features covering gammatone-domain, autocorrelation-domain, and modulation-domain features, as well as linear prediction features, MFCC variants, pitch-based features, etc, in various adverse acoustic environments. Because STFT has a perfect inverse transform, the log spectral magnitude becomes popular \citep{xu2015regression}. Recently, \cite{delfarah2017features} performed another feature
study in room reverberant situations, where log spectral magnitude and log mel-spectrum features were further added to the comparison. The conclusions in \citep{chen2014feature} and \citep{delfarah2017features} are consistent. Although learnable features are becoming a new trend \citep{tan2018gated,luo2019conv,pandey2019new}, very recent research results demonstrate that handcrafted acoustic features are still competitive to the learnable filters, e.g. \citep{ditter2020multi,pariente2020filterbank}. For multichannel speech enhancement, interaural time difference, interaural level difference \citep{jiang2014binaural}, interaural phase difference, and their variants \citep{yang2019boosting} are widely used spatial features. See \citep[Section 4]{wang2018supervised} for an excellent summary on the acoustic features.

 \section{Notations}\label{sec:notations}
We first introduce some notations here. Regular lower-case letters, e.g. $s$, $f$, and $\gamma$, indicate scalars. Bold lower-case letters, e.g. $\y$ and $\bm\alpha$, indicate vectors. Bold capital letters, e.g. $\mathbf{P}$ and $\bm\Phi$, indicate matrices. Letters in calligraphic fonts, e.g. $\mathcal{X}$, indicate sets. $\mathbf{0}$ ($\mathbf{1}$) is a vector with all entries being 1 (0). The operator $^T$ denotes the transpose. The operator $^H$ denotes the conjugate transpose of complex numbers.

  \begin{figure*}[!t]
\centering
\includegraphics[width=17.3cm]{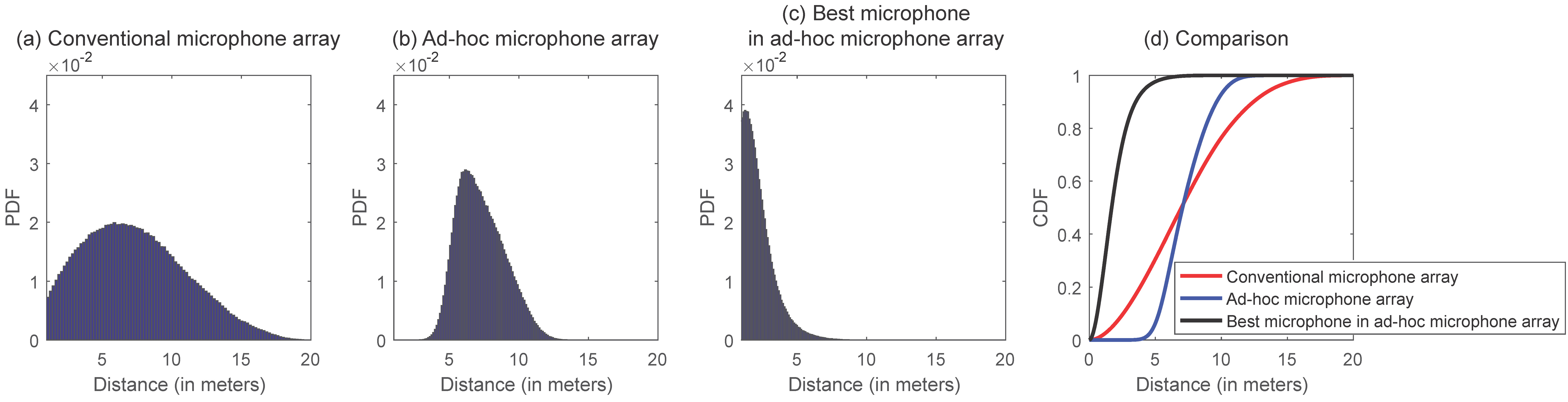}
\caption{Monte Carlo simulation of the distance distribution between a speech source and a microphone array in comparison. The physical spaces for this simulation contain a square room, a rectangle room, and a circle room (see sFig. 1 in the supplementary materials for the details of the three rooms). The farthest distance between the speech source and the microphone array in any of the rooms is set to 20 meters. Each microphone array in comparison consists of 16 microphones. (a) Probability density function (PDF) of the distance distribution of a conventional microphone array. The mean and standard deviation of this distribution are 7.28 and 3.71 meters respectively. (b) PDF of the distance distribution of an ad-hoc microphone array, where the distance is defined as the average distance between the speaker and each microphone in the ad-hoc array. The mean and standard deviation of this distribution are 7.28 and 1.68 meters respectively. (c) PDF of the distribution of the distance between the speech source and the best microphone in the ad-hoc microphone array, where the word ``best microphone'' denotes the closest microphone to the speech source.  The mean and standard deviation of the distribution are 1.92 and 1.21 meters respectively. (d) Cumulative distribution functions (CDF) of the distance distributions in Figs. \ref{fig:distribution_multi}a, \ref{fig:distribution_multi}b, and \ref{fig:distribution_multi}c.} \label{fig:distribution_multi}
\end{figure*}

 \begin{figure*}[!t]
\centering
\includegraphics[width=18cm]{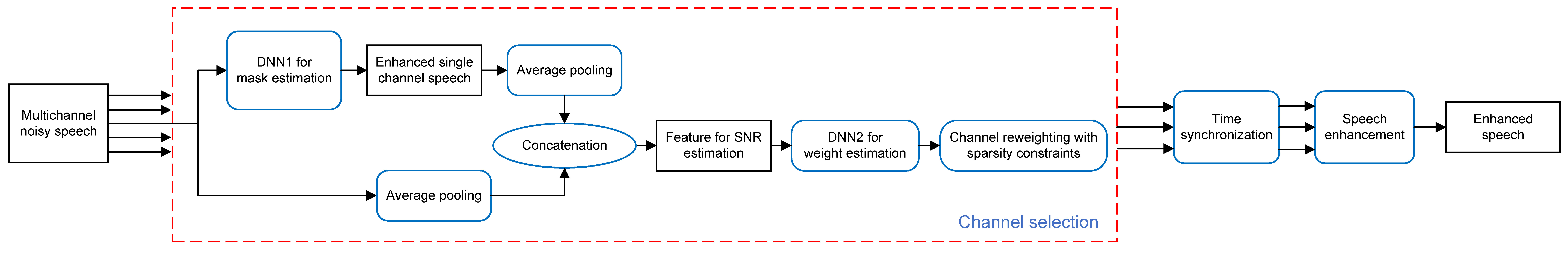}
\caption{Diagram of deep ad-hoc beamforming. The channel-selection framework is described in the red dashed box.} \label{fig:dab}
\end{figure*}

\section{Signal model of ad-hoc microphone arrays}\label{sec:signal_model}

 Ad-hoc microphone arrays can significantly reduce the probability of the occurrence of far-field environments. We take the case described in Fig.  \ref{fig:distribution_multi} as an example. When a speaker and a microphone array are distributed randomly in a room, the distribution of the distance between the speaker and an ad-hoc microphone array has a smaller variance than that between the speaker and a conventional microphone array (Figs. \ref{fig:distribution_multi}a and \ref{fig:distribution_multi}b). For example, the conventional array has a probability of 24\% to be placed over 10 meters away from the speech source, while the number regarding to the ad-hoc array is only 7\%. Particularly, the distance between the best microphone in the ad-hoc array and the speech source is only 1.9 meters on average, and the probability of the distance that is larger than 5 meters is only 2\% (Fig. \ref{fig:distribution_multi}c). 

Here we build the signal model of an ad-hoc microphone array.
All speech enhancement methods throughout the paper operate in the frequency domain on a frame-by-frame basis.
Suppose that a physical space contains one target speaker, multiple noise sources, and an ad-hoc microphone array of $M$ microphones. The physical model for the signals arrived at the ad-hoc array is assumed to be
\begin{eqnarray}\label{eq:1}
  \mathbf{v}(t,f) = \mathbf{c}(f)s(t,f) +\h(t,f)+ \mathbf{n}(t,f)
\end{eqnarray}
where $s(t,f)$ is the short-time Fourier transform (STFT) value of the target clean speech at time $t$ and frequency $f$, $\c(f)$ is the time-invariant acoustic transfer function from the speech source to the array which is an $M$-dimensional complex number:
\begin{eqnarray}
  \mathbf{c}(f) = [c_1(f),c_2(f),\ldots, c_M(f)]^T
\end{eqnarray}
$\mathbf{c}(f)s(t,f)$ and $\h(t,f)$ are
the direct sound and early and late reverberation of the target signal, and $\n(t,f)$ is the additive noise:
\begin{eqnarray}
  &&\mathbf{n}(t,f) = [n_1(t,f),n_2(t,f),\ldots, n_M(t,f)]^T\\
  &&\mathbf{v}(t,f) = [v_1(t,f),v_2(t,f),\ldots, v_M(t,f)]^T.
\end{eqnarray}
which are the STFT values of the received signals by the $m$-th microphone at time $t$ and frequency $f$.  Usually, we denote $\x(t,f) =  \mathbf{c}(f)s(t,f)$.

After processed by the devices $\{D_m(\cdot)\}_{m=1}^M$ where the microphones are fixed, the signals that the DAB finally receives are:
\begin{eqnarray}\label{eq:u}
  z_m(t,f) = D_m({v}_m(t,f)),\quad \forall m = 1,\ldots,M
\end{eqnarray}
with $\mathbf{z}(t,f) = [z_1(t,f),\ldots,z_M(t,f)]^T$.
Real-world devices $\{D_m(\cdot)\}_{m=1}^M$ may cause many problems including the unsynchronization of time delay, clock rates, adaptive gain controllers, etc. Here we consider the time unsynchronization problem:
\begin{equation}\label{eq:v}
\begin{aligned}
  z_m(t,f) &= {v}_m(t+\tau_m,f)\\
  &={x_m}(t+\tau_m,f) +h_m(t+\tau_m,f)+ {n}_m(t+\tau_m,f)
\end{aligned}
\end{equation}
where $\tau_m$ is the time delay caused by the $m$th device.

\section{Deep ad-hoc beamforming: A system overview}\label{sec:db_ama}

 A system overview of DAB is shown in Fig. \ref{fig:dab}. It contains three core components---a supervised channel selection framework, a supervised time synchronization framework, and a speech enhancement module.

  The core idea of the channel selection framework is to filter the received signals $\z(t,f)$ by a channel-selection vector $\mathbf{p}=[p_1,\ldots,p_M]^T$:
\begin{equation}\label{eq:4}
  \mathbf{z}_{\p}(t,f) = \p\circ\mathbf{z}(t,f)
\end{equation}
such that the channels that output low quality speech signals can be suppressed or even discarded,
where $\mathbf{p}$ is the output mask of the channel-selection method described in the red box of Fig. \ref{fig:dab}, and $\circ$ denotes the element-wise product operator. Without loss of generality, we assume the selected channels are ${z}_{1}(t,f),\ldots, z_N(t,f)$.

The time synchronization module first selects the noisy signal from the best channel, assumed to be $z_k(t,f)$, as a reference signal by a supervised 1-best channel selection algorithm that will be described in Section \ref{sec:dab_3}. Then, it estimates the relative time delay of the noisy signals from the selected microphones over the reference signal by a time delay estimator:
 \begin{equation}\label{eq:time_delay_estimation}
  \hat{\tau}_n = h(z_n(t,f)|z_k(t,f)),\quad\forall n = 1,\ldots,N
\end{equation}
where $h(z_n(t,f)|z_k(t,f))$ is the time delay estimator with $z_k(t,f)$ as the reference signal, and $\hat{\tau}_n$ is the estimated relative time delay of $z_n(t,f)$ over $z_k(t,f)$.
Finally, it synchronizes the microphones according to the estimated time delay:
  \begin{equation}\label{eq:synchronization}
  {y}_n(t,f) = z_n(t-\hat{\tau}_n,f),\quad\forall n = 1,\ldots,N
\end{equation}
Note that $\hat{\tau}_n$ consists of the relative time delay caused by both the device and the signal transmission through air.
Because developing a new accurate time delay estimator is not the focus of this paper, we simply use the classic generalized cross-correlation phase transform \citep{knapp1976generalized,carter1987coherence} as the estimator, though many other time delay estimators can be adopted as well \citep{chen2006time}. For example, the deep neural network based time delay estimators \citep{wang2018robust}, which were original proposed for the estimation of the signal direction of arrival, can be adopted here too.

The speech enhancement module takes $\y(t,f) = [y_1(t,f),\ldots,y_N(t,f)]^T $ as its input. Many deep-learning-based speech enhancement methods can be used directly or with slight modification as the speech enhancement module. Here we take the MVDR based deep beamforming as an example. Because the deep beamforming is trained in a single-channel fashion as the other parts of DAB, it makes the overall DAB flexible in incorporating any number of microphones in the test stage without retraining or modifying the DAB model. This is an important requirement of real world applications that should also be considered in other DAB implementations.
Note that, if $N=1$, then DAB outputs the noisy speech of the selected single channel directly without resorting to deep beamforming anymore.

In the following two sections, we will present the supervised channel selection framework and speech enhancement module respectively.


\section{Supervised channel selection}\label{subsec:cs}
The channel-selection algorithm is applied to each channel \textit{independently}. It contains two steps described in the following two subsections respectively.

\subsection{Channel-reweighting model}\label{sec:dab_2}

Suppose there is a test utterance of $U$ frames, and suppose the received speech signal at the $i$-th channel is $\{\widetilde{\z}_i(t)\}_{t=1}^U$:
 \begin{eqnarray}\label{eq:10}
\widetilde{\z}_i(t) = [{|z|}_i(t,1),\ldots, {|z|}_i(t,F)]^T
\end{eqnarray}
where ${|z|}_i(t,f)$ is the amplitude spectrogram of $\z(t,f)$ at the $i$-th channel.

The channel-reweighting model estimates the channel weight $q_i$ of the $i$th channel by
  \begin{eqnarray}\label{eq:core}
{q}_i = g\left(\bar{\widetilde{\z}}_i\right)
\end{eqnarray}
where $g(\cdot)$ is a DNN-based channel-reweighting model, and $\bar{\widetilde{\z}}_i$ is the average pooling result of $\{\widetilde{\z}_i(t)\}_{t=1}^U$:
 \begin{eqnarray}\label{eq:channel_weight}
 \bar{\widetilde{\z}}_i = \frac{1}{U}\sum_{t=1}^U \widetilde{\z}_i(t)
\end{eqnarray}
 To train $g(\cdot)$, we need to first define a training target, and then extract noise-robust handcrafted features, which are described as follows.

\subsubsection{Training targets}

  This paper uses a variant of SNR as the target:
  \begin{eqnarray}\label{eq:x}
\frac{ \sum_t|x|_{\rm{time}}(t)}{ \sum_t|x|_{\rm{time}}(t) +\sum_t|n|_{\rm{time}}(t)}
\end{eqnarray}
where $\{x_{\rm{time}}(t)\}_t$ and $\{n_{\rm{time}}(t)\}_t$ are the direct sound and additive noise of the received noisy speech signal in time-domain.

Many measurements may be used as the training targets as well, such as the equivalent form of \eqref{eq:x} in the time-frequency domain
 \begin{eqnarray}\label{1..1}
\frac{\sum_{t=1}^U\sum_{f=1}^F|x(t,f)|}{\sum_{t=1}^U\sum_{f=1}^F|x(t,f)|+\sum_{t=1}^U\sum_{f=1}^F|h(t,f)+n(t,f)|},
\end{eqnarray}
performance evaluation metrics including signal-to-distortion ratio (SDR), short-time objective intelligibility (STOI) \citep{taal2011algorithm}, etc., application driven metrics including equal error rate (EER) \citep{bai2020cosine}, partial area under the ROC curve \citep{bai2020speaker,bai2020partial}, word error rate, etc, as well as other device-specific metrics including the battery life of a cell phone, etc. For example, if a cell phone is to be out of power, then DAB should prevent the cell phone being an activated channel.

\subsubsection{Handcrafted features}

 Because the STFT feature $\{\widetilde{\z}_i(t)\}_{t=1}^U$ may not be noise-robust enough, an important issue to the effectiveness of \eqref{eq:core} is the acoustic feature. Here we introduce two handcrafted features---enhanced STFT (eSTFT) and multi-resolution cochleagram (MRCG) feature.

\textit{1) eSTFT:} As shown in the red dashed box of Fig. \ref{fig:dab}, we first use a DNN-based single-channel speech enhancement method, denoted as DNN1, to generate an estimated \textit{ideal ratio mask} (IRM) of the direct sound of $\widetilde{\z}_i(t)$, denoted as $\{\hat{\x}_i(t)\}_{t=1}^U$:
 \begin{eqnarray}\label{eq:100}
\hat{\x}_i(t) = [\widehat{\rm{\textit{IRM}}}_i(t,1),\ldots, \widehat{\rm{\textit{IRM}}}_i(t,F)]^T
\end{eqnarray}
where $\widehat{\rm{\textit{IRM}}}_i(t,f)$ is the estimate of the IRM at the $i$-th channel. The IRM is the training target of DNN1:
 \begin{equation}\label{1..1}
{\rm{\textit{IRM}}}(t,f) = \frac{|x(t,f)|}{|x(t,f)|+|h(t,f)+n(t,f)|}
\end{equation}
 where $|x(t,f)|$, $|h(t,f)|$, and $|n(t,f)|$ are the amplitude spectrograms of the direct and early reverberant speech, late reverberant speech, and noise components of single-channel noisy speech respectively. Then, we denote the concatenation of the estimated IRM $\hat{\x}_i(t)$ and the noisy feature $\widetilde{\z}_i(t)$ as the \textit{eSTFT} feature, which is used to replace $\widetilde{\z}_i(t)$ in \eqref{eq:channel_weight}.

 To discriminate the channel selection model $g(\cdot)$ from DNN1, we denote $g(\cdot)$ as DNN2.
 As presented above, both DNN1 and DNN2 are trained on single-channel data only instead of multichannel data from ad-hoc microphone arrays, which is an important merit for the practical use of DAB. In practice, the training data of DNN1 and DNN2 need to be independent so as to prevent overfitting.

\textit{2) MRCG:} Another alternative of STFT can be MRCG, which has shown to be a noise robust acoustic feature for speech separation \citep{chen2014feature}.\footnote{Code is downloadable from http://web.cse.ohio-state.edu/pnl/software.html}
The key idea of MRCG is to incorporate both global information and local information of speech through multi-resolution extraction. The global information is produced by extracting cochleagram features with a large frame length or a large smoothing window (i.e., low resolutions). The local information is produced by extracting cochleagram features with a small frame length and a small smoothing window (i.e., high resolutions). It has been shown that cochleagram features with a low resolution, such as frame length = 200 ms, can detect patterns of noisy speech better than that with only a high resolution, and features with high resolutions complement those with low resolutions. Therefore, concatenating them together is better than using them separately. In this paper, we adopt the implementation in \citep{zhang2016boosting}.

Because $g(\cdot)$ does not need to recover the time-domain signal, many handcrafted acoustic features can be used beyond the above two examples to further improve the estimation accuracy. Some candidate acoustic features are listed in \citep{chen2014feature,guido2018tutorial}. Besides, a large family of wavelet transforms \citep{mallat1989theory} have not been deeply studied yet. Here we list some possible candidates for DAB: wavelet-packet transform \citep{sepulveda2013estimation}, discrete shapelet transform \citep{guido2018fusing}, fractal-wavelet transform \citep{guariglia2016fractional,guariglia2019primality}, and adaptive multiscale wavelet transform \citep{zheng2019framework}.

\subsection{Channel-selection algorithms}\label{sec:dab_3}
Given the estimated weights $\mathbf{q}=[q_1,\ldots,q_M]^T$ of the test utterance, many advanced sparse learning methods are able to project $\mathbf{q}$ to $\p$, i.e. $\mathbf{p} = \delta(\mathbf{q})$
 where $\delta(\cdot)$ is a channel-selection function that enforces sparse constraints on $\mathbf{q}$. This section designs several $\delta(\cdot)$ functions as follows.

\subsubsection{One-best channel selection (1-best)}
The simplest channel-selection method is to pick the channel with the highest SNR:
  \begin{eqnarray}
{p}_i = \left\{\begin{array}{ll}
  1,& \quad \mbox{if } q_i = \max_{1\leq k \leq M} q_k\\
  0,& \quad \mbox{otherwise}
\end{array}\quad \forall i = 1,\ldots,M.\right.
\end{eqnarray}
After the channel selection, DAB outputs the noisy speech from the selected channel directly.

\subsubsection{All channels (all-channels)}
Another simple channel-selection method is to select all channels with equivalent importance:
  \begin{eqnarray}
{p}_i =  1,\quad \forall i = 1,\ldots,M.
\end{eqnarray}
This method is an extreme case of channel selection that usually performs well when the microphones are distributed in a small space.

\subsubsection{\textit{N}-best channel selection with predefined $N$ (fixed-N-best)}
When the microphone number $M$ is large enough, there might exist several microphones close to the speech source whose received signals are more informative than the others. It is better to group the informative microphones together into a local array instead of selecting one best channel:
  \begin{eqnarray}\label{eq:nbest}
{p}_i = \left\{\begin{array}{ll}
  1,& \quad \mbox{if } q_i \in \{q'_1,q'_2,\ldots,q'_N\} \\
  0,& \quad \mbox{otherwise}
\end{array}\quad \forall i = 1,\ldots,M.\right.
\end{eqnarray}
where $q'_1\geq q'_2\geq\ldots\geq q'_M$ is the descent order of $\{q_i\}_{i=1}^M$, and $N$ is a user-defined hyperparameter, $N\leq M$.

\subsubsection{\textit{N}-best channel selection where $N$ is determined on-the-fly (auto-N-best)}

Here we develop a simple method that determines the hyperparameter $N$ in \eqref{eq:nbest} on-the-fly. It first finds $q_{*} = \max_{i\in\{1,\ldots,M\}}q_i$, and then determines $\mathbf{p}$ by
  \begin{equation}\label{14}
{p}_i = \left\{\begin{array}{ll}
 1,&\quad \mbox{if } \frac{q_{i}}{q_{*}}\frac{1-q_{*}}{1-q_i}>\gamma\\
 0,&\quad \mbox{otherwise}
\end{array}\right.,\quad\forall i = 1,\ldots,M.
\end{equation}
where $\gamma\in[0,1]$ is a tunable threshold. See Appendix for the proof of \eqref{14}.

\subsubsection{Soft N-best channel selection (soft-N-best)}
One way to encode the signal quality of the selected channels in \eqref{14} is to use soft weights as follows:
  \begin{equation}\label{15}
{p}_i = \left\{\begin{array}{ll}
 q_i,&\quad \mbox{if } \frac{q_{i}}{q_{*}}\frac{1-q_{*}}{1-q_i}>\gamma\\
 0,&\quad \mbox{otherwise}
\end{array}\right.,\quad\forall i = 1,\ldots,M.
\end{equation}

\subsubsection{Machine-learning-based N-best channel selection (learning-N-best)}

\begin{figure*}[!t]
\centering
\includegraphics[width=15cm]{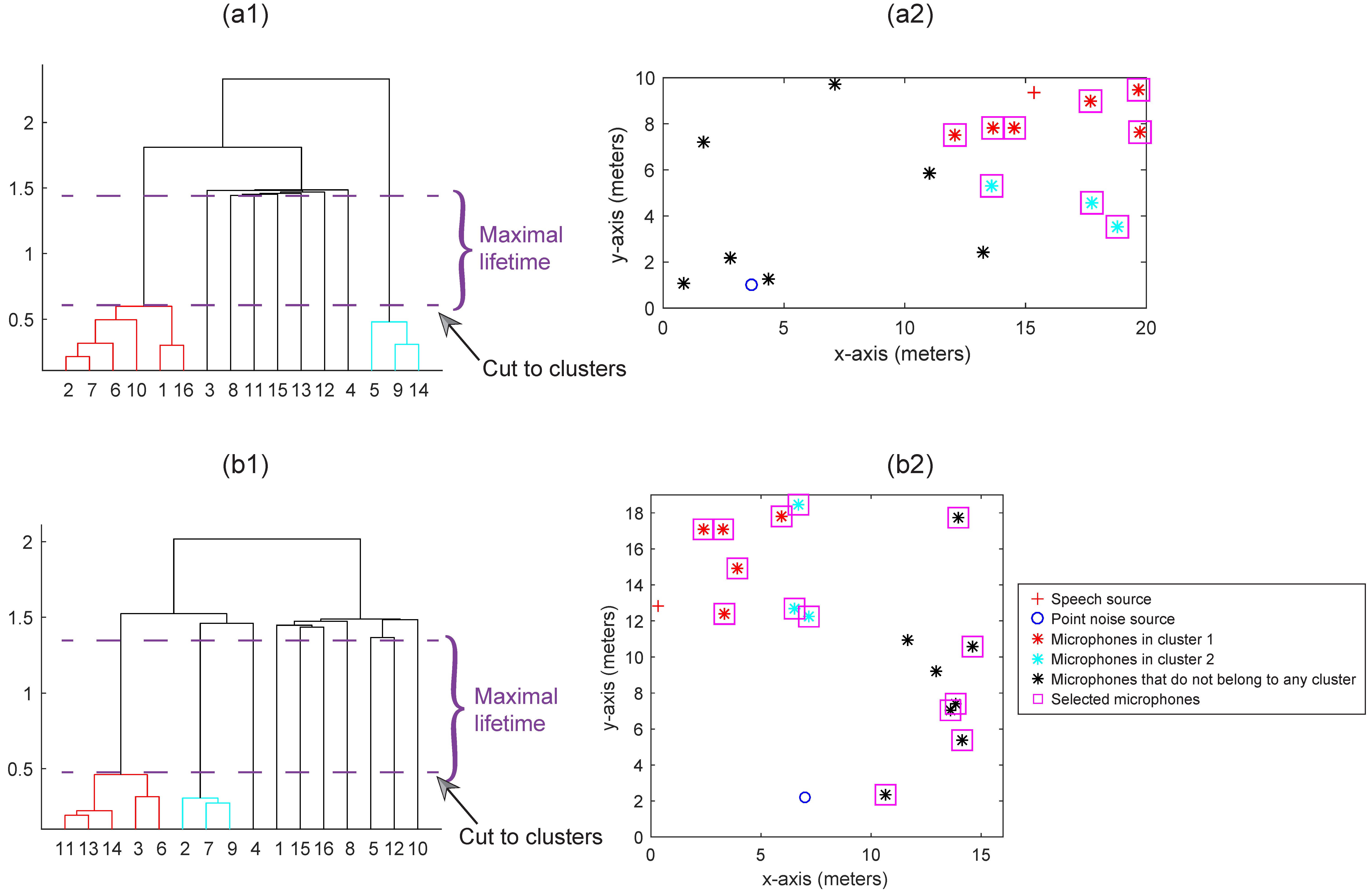}
\caption{Two examples of the learning-\textit{N}-best channel selection method in point source noise environments. (a1) Dendrogram of Example 1. (b1) Channel selection result of Example 1. (a2) Dendrogram of Example 2. (b2) Channel selection result of Example 2.} \label{fig:channel_learning_N_best}
\end{figure*}

The above channel selection methods determine the selected channels by SNR only, without considering the correlation between the channels. As we know, the correlation between the channels, which encodes environmental information and the time delay between the microphones, is important to adaptive beamforming. Here, we develop a spectral clustering based channel selection method that takes the correlation into the design of the affinity matrix of the spectral clustering.

Unlike the other channel selection algorithms, ``learning-\textit{N}-best'' should first conduct the time synchronization, which takes $\y(t,f) = [y_1(t,f),\ldots,y_M(t,f)]^T$ as its input.
Then, it calculates the covariance matrix of the noisy speech across the channels by
  \begin{equation}\label{kernel}
\bm\Phi_{\y\y}(f) = \sum_{t} \y(t,f)\y(t,f)^H.
\end{equation}
and normalize \eqref{kernel} to an amplitude covariance matrix $\bm\Phi_{\y\y}^{\rm{norm}}(f)$:
 \begin{eqnarray}\label{kernel2}
\Phi_{\y\y}^{\rm{norm}}(f)(i,j) = \frac{|\Phi_{\y\y}(f)(i,j)|^2}{\Phi_{\y\y}(f)(i,i)\Phi_{\y\y}(f)(j,j)},\nonumber\\
\quad \forall i,j =1,\ldots,M.
\end{eqnarray}
After that, it calculates a new matrix $\mathbf{K}$ by averaging the amplitude covariance matrix along the frequency axis by
 \begin{eqnarray}\label{kernel2}
K(i,j) = \frac{1}{F}\sum_{f=1}^F \Phi_{\y\y}^{\rm{norm}}(f)(i,j),
\quad \forall i,j =1,\ldots,M
\end{eqnarray}
where $F$ is the number of the DFT bins. The affinity matrix $\mathbf{A}$ of the spectral clustering is defined as
 \begin{eqnarray}\label{kernel2}
\mathbf{A} = \exp\left(-\frac{|\K-\mathbf{I}|^2}{2\sigma^2}\right),
\quad \forall i,j =1,\ldots,M
\end{eqnarray}
where $\I$ is the identity matrix, and $\sigma$ is a hyperparameter with a default value $1$. Following the Laplacian eigenvalue decomposition \citep{ng2001spectral} of $\mathbf{A}$, it obtains a $J\times M$-dimensional representation of the channels, $\mathbf{U}=[\u_1,\ldots,\u_M]$, where $\u_i$ is the representation of the $i$th microphone and $J$ denotes the dimension of the representation.

``Learning-\textit{N}-best'' conducts agglomerative hierarchical clustering on $\mathbf{U}$, and takes the maximal lifetime of the dendrogram as the threshold to partition the microphones into $B$ clusters ($1\leq B\leq M$), denoted as $\mathcal{U}_1,\ldots,\mathcal{U}_B$. The maximum predicted SNRs of the microphones in the clusters are denoted as $q'_1,\ldots,q'_B$ respectively. Finally, it groups the microphones that satisfy the following condition into a local microphone array:
 \begin{eqnarray}\label{eq:leanring}
 {p}_i = \left\{\begin{array}{ll}
 1,&\quad \mbox{if } \u_i\in \mathcal{U}_b \mbox{ and } \frac{q'_{b}}{q'_{*}}\frac{1-q'_{*}}{1-q'_b}>\gamma\\
 0,&\quad \mbox{otherwise}
\end{array}\right.,\nonumber\\
\quad\forall i = 1,\ldots, M,\quad\forall b= 1,\ldots,B.
\end{eqnarray}
where $q'_*= \max_{1\leq b\leq B} q'_b$.

Figure \ref{fig:channel_learning_N_best} lists two examples of the ``learning-\textit{N}-best'' method. From the figure, we see that the microphones around the speech sources are grouped into clusters, while the microphones that are far away from the speech sources have weak correlations, hence they form a number of special clusters that each contains only one microphone. Note that, as shown in Example 2 of Fig. \ref{fig:channel_learning_N_best}, because the selection criterion is determined by \eqref{eq:leanring}, there is no guarantee that the clusters that contain more than one microphone will be selected, or the clusters that contain only a single microphone will be discarded.


\section{Speech enhancement: An application case}\label{sec:problem}


After getting the synchronized signals $\y(t,f) = [y_1(t,f),\ldots,y_N(t,f)]^T $, we may use existing multichannel signal processing techniques directly or with slight modification for a specific application. Here we use a deep beamforming algorithm \citep{heymann2016neural,wang2018all} directly for speech enhancement as an example.

The deep beamforming algorithm finds a linear estimator $\w_{\rm{opt}}(f)$ to filter $\mathbf{y}(t,f)$ by the following equation:
\begin{eqnarray}\label{eq:3}
  \hat{x}_{\rm{ref.}}(t,f) = \w_{\rm{opt}}^H(f)\mathbf{y}(t,f).
\end{eqnarray}
where $\hat{x}_{\rm{ref.}}(t,f)$ is an estimate of the direct sound at the reference microphone of the array. For example, MVDR finds $\w_{\rm{opt}}$ by minimizing the average output power of the beamformer while maintaining the energy along the target direction:
\begin{eqnarray}\label{eq:general}
 \min_{\w(f)}&\mbox{ }\w^H(f)\bm\Phi_{\n\n}(f)\w(f)\\
  {\rm{subject \mbox{ }to}}&\w^H(f)\c(f) = 1\nonumber
\end{eqnarray}
where $\bm\Phi_{\n\n}(f)$ is an $M\times M$-dimensional cross-channel covariance matrix of the received noise signal $\n(f)$. \eqref{eq:general} has a closed-form solution:
 \begin{eqnarray}\label{eq:6}
  \w_{\rm{opt}}(f) &=& \frac{\widehat{\bm\Phi}^{-1}_{\n\n}(f)\hat{\mathbf{c}}(f)}{\hat{\mathbf{c}}^H(f)\widehat{\bm\Phi}^{-1}_{\n\n}(f)\hat{\mathbf{c}}(f)}
\end{eqnarray}
where the variables $\widehat{\bm\Phi}_{\n\n}(f)$ and $\hat{\c}(f)$ are the estimates of ${\bm\Phi}_{\n\n}(f)$ and ${\c}(f)$ respectively which are derived by the following equations according to \citep{zhang2017speech,wang2018all}:
 \begin{eqnarray}\label{eq:y}
 &&\widehat{\bm\Phi}_{\x\x}(f)=\frac{1}{\sum_t \eta(t,f)}\sum_t \eta(t,f)\y(t,f)\y(t,f)^H\\
&&\widehat{\bm\Phi}_{\n\n}(f)=\frac{1}{\sum_t \xi(t,f)}\sum_t \xi(t,f)\y(t,f)\y(t,f)^H\\
&&\hat{\c}(f) = {\rm{principal}}\left( \widehat{\bm\Phi}_{\x\x}(f) \right)
\end{eqnarray}
where $\widehat{\bm\Phi}_{\x\x}(f)$ is an estimate of the covariance matrix of the direct sound $\x(t,f)$, ${\rm{principal}}(\cdot)$ is a function returning the first principal component of the input square matrix, and $\eta(t,f)$ and $\xi(t,f)$ are defined as the product of individual estimated T-F masks:
 \begin{eqnarray}\label{eq:y}
&&\eta(t,f) = \prod_{i=1}^M \widehat{\rm{\textit{IRM}}}_i(t,f)\\
&&\xi(t,f) = \prod_{i=1}^M \left(1-\widehat{\rm{\textit{IRM}}}_i(t,f)\right)
\end{eqnarray}
Note that, in our experiments, when we calculate $\eta(t,f)$ and $\xi(t,f)$, we take all channels of the ad-hoc array into consideration, which empirically results in slight performance improvement over the method that we take only the selected channels into the calculation.

\section{Experiments}\label{sec:no_delay}

In this section, we study the effectiveness of DAB in diffuse noise and point source noise environments under the situation where the output signals of the channels have random time delay caused by devices. Specifically, we first present the experimental settings in Section \ref{subsec:exp_set}, then present the experimental results in the diffuse noise and point source noise environments in Section \ref{sec:match}, and finally discuss the effects of hyperparameter settings on performance in Sections \ref{subsec:number_phone} and \ref{subsec2:hyper}.

\subsection{Experimental settings}\label{subsec:exp_set}

\textit{Datasets:} The clean speech was generated from the TIMIT corpus. We randomly selected half of the training speakers to construct the database for training DNN1, and the remaining half for training DNN2. We used all test speakers for test. The noise source for the training database was a large-scale sound effect library which contains over 20,000 sound effects. The additive noise for the test database was the babble, factory1, and volvo noise respectively from the NOISEX-92 database.

\textit{Training data:} We simulated a rectangle room for each training utterance. The length and width of the rectangle room were generated randomly from a range of $[5,30]$ meters. The height was generated randomly from $[2.5, 4]$ meters. The reverberant environment was simulated by an image-source model.\footnote{https://github.com/ehabets/RIR-Generator} Its T60 was selected randomly from a range of $[0, 1$] second.
A speech source, a noise source, and a single microphone were placed randomly in the room. The SNR, which is the energy ratio between the speech and noise at the locations of their sources, was randomly selected from a range of $[-10,20]$ dB. We synthesized 50,000 noisy utterances to train DNN1, and 100,000 noisy utterances to train DNN2.

\textit{Test data:} We constructed a rectangle room for each test utterance. The length, width, and height of the room were randomly generated from $[10, 20]$, $[10,20]$, and $[2.7, 3.5]$ meters respectively. The additive noise is assumed to be either diffuse noise or point source noise.

For the diffuse noise environment, a speech source and a microphone array were placed randomly in the room. The T60 for generating reverberant speech was selected randomly from a range of $[0.4, 0.8$] second. To simulate the uncorrelated diffuse noise, the noise segments at different microphones do not have overlap, and they were added directly to the reverberant speech at the microphone receivers without referring to reverberation. The noise power at the locations of all microphones were maintained at the same level, which was calculated by the SNR of the direct sound over the additive noise at a place of 1 meter away from the speech source, denoted as $\textit{the SNR at the origin}$ (SNRatO). Note that the SNRs at different microphones were different due to the energy degradation of the speech signal during its propagation. The SNRatO was selected from 10, 15, and 20 dB respectively. We generated 1,000 test utterances for each SNRatO, each noise type, and each kind of microphone array, which amounts to 9 test scenarios and 18,000 test utterances.

For the point source noise environment, a speech source, a point noise source, and a microphone array were placed randomly in the room. The T60 of the room was selected randomly from a range of $[0.4, 0.8$] second for generating reverberant speech and reverberant noise at the microphone receivers. The SNRatO was defined as the log ratio of the speech power over the noise power at their source locations respectively. It was chosen from $\{-5,5,15\}$ dB. Like the diffuse noise environment, we also generated 1,000 test utterances for each SNRatO, each noise type, and each kind of microphone array.

For both of the test environments, we generated a random time delay $\tau$ from a range of $[0, 0.5]$ second at each microphone of an ad-hoc microphone array for simulating the time delay caused by devices.

\textit{Comparison methods:} The baseline is the MVDR-based DB \citep{heymann2016neural} with a linear array of 16 microphones, which is described in Section \ref{sec:problem}. All DB models employed DNN1 for the single-channel noise estimation. The aperture size of the linear microphone array (i.e. the distance between two neighboring microphones) was set to 10 centimeters.

 DAB also employed an ad-hoc array of 16 microphones. We denote the DAB with different channel selection algorithms as:
\begin{itemize}
  \item \textbf{DAB+1-best.}
  \item \textbf{DAB+all-channels.}
  \item \textbf{DAB+fixed-\textit{N}-best.} We set $N=\sqrt{M}$.
  \item \textbf{DAB+auto-\textit{N}-best.} We set $\gamma =0.5$.
  \item \textbf{DAB+soft-\textit{N}-best.} We set $\gamma =0.5$.
  \item \textbf{DAB+learning-\textit{N}-best.} We set $J=M/2$, $\sigma=1$, and $\gamma =0.5$.
\end{itemize}
To study the effectiveness of time synchronization (TS) module, we further compared the following systems:
\begin{itemize}
  \item \textbf{DAB+channel selection method.} It does not use the TS module.
  \item \textbf{DAB+channel selection method+GT.} It uses the ground truth (GT) time delay caused by the devices to synchronize the microphones.
  \item \textbf{DAB+channel selection method+TS.} It uses the TS module to estimate the time delay caused by both different locations of the microphones and different devices where the microphones are installed.
\end{itemize}
We implemented the above comparison methods with different channel selection algorithms.
For example, if the channel selection algorithm is ``auto-\textit{N}-best'', then the comparison systems are ``DAB+auto-\textit{N}-best'', ``DAB+auto-\textit{N}-best+GT'', and ``DAB+auto-\textit{N}-best+TS'' respectively.

\textit{DNN models:} For each comparison method, we set the frame length and frame shift to 32 and 16 milliseconds respectively, and extracted 257-dimensional STFT features. We used the same DNN1 for DB and DAB. DNN1 is a standard feedforward DNN. It contains two hidden layers. Each hidden layer has 1024 hidden units. The activation functions of the hidden units and output units are rectified linear unit and sigmoid function, respectively. The number of epochs was set to 50. The batch size was
set to 512. The scaling factor for the adaptive stochastic gradient
descent was set to 0.0015, and the learning rate decreased linearly
from 0.08 to 0.001. The momentum of the first 5 epochs
was set to 0.5, and the momentum of other epochs was set to
0.9. A contextual window was used to expand each input frame to its context along the time axis. The window size was set to 7.

DNN2 has the same parameter setting as DNN1 except that DNN2 does not need a contextual window, was trained with a batch size of 32, and took eSTFT as the acoustic feature. All DNNs were well-tuned. Note that although \textit{bi-directional long short-term memory} may lead to better performance, we simply used the feedforward DNN since the type of the DNN models is not the focus of this paper.

\textit{Evaluation metrics:} The performance evaluation metrics include STOI \citep{taal2011algorithm}, perceptual evaluation of speech quality (PESQ) \citep{rix2001perceptual}, and signal to distortion ratio (SDR) \citep{vincent2006performance}. STOI evaluates the objective speech intelligibility of time-domain signals. It has been shown empirically that STOI scores are well correlated with human speech intelligibility scores \citep{wang2014training,du2014speech,huang2015joint,zhang2016deep}. PESQ is a test methodology for automated assessment of the speech quality as experienced by a listener of a telephony system. SDR is a metric similar to SNR for evaluating the quality of enhancement. The higher the value of an evaluation metric is, the better the performance is.



\subsection{Main results}\label{sec:match}

\begin{table}[!t]
  \centering
  \caption{Results with 16 microphones per array in diffuse noise environments.}
  \scalebox{0.5}{
    \begin{tabular}{|r|l|rrr|rrr|rrr|}
    \hline
    \multicolumn{1}{|r|}{\multirow{2}[4]{*}{SNRatO}} & \multirow{2}[4]{*}{Comparison methods} & \multicolumn{3}{c|}{Babble} & \multicolumn{3}{c|}{Factory} & \multicolumn{3}{c|}{Volvo} \bigstrut\\
\cline{3-11}          &       & \multicolumn{1}{c}{STOI} & \multicolumn{1}{c}{PESQ} & \multicolumn{1}{c|}{SDR} & \multicolumn{1}{c}{STOI} & \multicolumn{1}{c}{PESQ} & \multicolumn{1}{c|}{SDR} & \multicolumn{1}{c}{STOI} & \multicolumn{1}{c}{PESQ} & \multicolumn{1}{c|}{SDR} \bigstrut\\
    \hline
    \multicolumn{1}{|r|}{\multirow{18}[12]{*}{10 dB}} & Noisy & 0.5989  & 1.86  & 1.12  & 0.5969  & 1.80  & 1.20  & 0.6785  & 2.10  & 1.62  \bigstrut[t]\\
          & DB    & 0.6911  & 1.87  & 2.75  & 0.6900  & 1.86  & 3.42  & 0.7766  & 2.16  & 3.95  \\
          & DAB (1-best) & 0.7154  & 2.06  & 5.14  & 0.7143  & 2.00  & 5.13  & 0.7892  & 2.31  & 5.23  \bigstrut[b]\\
\cline{2-11}          & DAB (all-channels) & 0.5824  & 1.83  & -0.93  & 0.5760  & 1.78  & -1.55  & 0.6061  & 1.88  & -1.49  \bigstrut[t]\\
          & DAB (all-channels+GT) & 0.7206  & 2.06  & 4.40  & 0.7137  & 2.00  & 4.47  & 0.7831  & 2.39  & 4.42  \\
          & DAB (all-channels+TS) & 0.7405  & 2.00  & 3.49  & 0.7388  & 1.95  & 3.54  & 0.8039  & 2.32  & 3.25  \bigstrut[b]\\
\cline{2-11}          & DAB (fixed-N-best) & 0.6026  & 1.87  & -0.12  & 0.6022  & 1.82  & -0.33  & 0.6351  & 1.92  & -0.32  \bigstrut[t]\\
          & DAB (fixed-N-best+GT) & 0.7451  & 2.12  & 5.10  & 0.7437  & 2.07  & 5.16  & 0.8117  & 2.42  & 5.54  \\
          & DAB (fixed-N-best+TS) & 0.7675  & 2.11  & 5.18  & 0.7634  & 2.05  & 5.01  & 0.8460  & 2.43  & 5.97  \bigstrut[b]\\
\cline{2-11}          & DAB (auto-N-best) & 0.5982  & 1.87  & -0.13  & 0.5927  & 1.83  & -0.61  & 0.6573  & 2.00  & 0.82  \bigstrut[t]\\
          & DAB (auto-N-best+GT) & 0.7531  & 2.14  & 5.74  & 0.7518  & 2.09  & 5.73  & 0.8164  & 2.46  & 5.97  \\
          & DAB (auto-N-best+TS) & 0.7696  & 2.12  & 5.45  & 0.7641  & 2.06  & 5.36  & 0.8405  & 2.44  & 5.85  \bigstrut[b]\\
\cline{2-11}          & DAB (soft-N-best) & 0.5999  & 1.85  & -0.28  & 0.5952  & 1.83  & -0.76  & 0.6645  & 2.01  & 0.84  \bigstrut[t]\\
          & DAB (soft-N-best+GT) & 0.7463  & 2.13  & 5.22  & 0.7455  & 2.07  & 5.26  & 0.8055  & 2.42  & 5.50  \\
          & DAB (soft-N-best+TS) & 0.7659  & 2.12  & 5.11  & 0.7610  & 2.05  & 5.09  & 0.8363  & 2.43  & 5.65  \bigstrut[b]\\
\cline{2-11}          & DAB (learning-N-best) & 0.5973  & 1.86  & -0.29  & 0.5892  & 1.81  & -0.99  & 0.6488  & 1.98  & 0.21  \bigstrut[t]\\
          & DAB (learning-N-best+GT) & 0.7405  & 2.12  & 5.22  & 0.7387  & 2.06  & 5.34  & 0.8026  & 2.43  & 5.35  \\
          & DAB (learning-N-best+TS) & 0.7631  & 2.07  & 4.55  & 0.7606  & 2.02  & 4.59  & 0.8330  & 2.41  & 4.97  \bigstrut[b]\\
    \hline
    \multicolumn{1}{|r|}{\multirow{18}[12]{*}{15 dB}} & Noisy & 0.6410  & 1.97  & 3.05  & 0.6400  & 1.93  & 2.79  & 0.6847  & 2.10  & 2.87  \bigstrut[t]\\
          & DB    & 0.7350  & 2.02  & 4.37  & 0.7396  & 1.99  & 4.61  & 0.7804  & 2.19  & 4.86  \\
          & DAB (1-best) & 0.7496  & 2.17  & 6.59  & 0.7527  & 2.14  & 6.45  & 0.7906  & 2.31  & 6.58  \bigstrut[b]\\
\cline{2-11}          & DAB (all-channels) & 0.5977  & 1.85  & -0.52  & 0.5990  & 1.84  & -0.87  & 0.6102  & 1.88  & -0.75  \bigstrut[t]\\
          & DAB (all-channels+GT) & 0.7575  & 2.22  & 5.45  & 0.7588  & 2.18  & 5.39  & 0.7887  & 2.42  & 5.23  \\
          & DAB (all-channels+TS) & 0.7809  & 2.15  & 4.53  & 0.7869  & 2.12  & 4.54  & 0.8091  & 2.35  & 4.32  \bigstrut[b]\\
\cline{2-11}          & DAB (fixed-N-best) & 0.6218  & 1.89  & 0.27  & 0.6189  & 1.88  & 0.03  & 0.6463  & 1.93  & 0.33  \bigstrut[t]\\
          & DAB (fixed-N-best+GT) & 0.7788  & 2.26  & 6.15  & 0.7832  & 2.22  & 6.11  & 0.8172  & 2.43  & 6.37  \\
          & DAB (fixed-N-best+TS) & 0.8074  & 2.26  & 6.65  & 0.8095  & 2.21  & 6.36  & 0.8518  & 2.44  & 7.04  \bigstrut[b]\\
\cline{2-11}          & DAB (auto-N-best) & 0.6188  & 1.90  & 0.44  & 0.6142  & 1.87  & -0.13  & 0.6641  & 2.00  & 1.44  \bigstrut[t]\\
          & DAB (auto-N-best+GT) & 0.7877  & 2.30  & 6.83  & 0.7946  & 2.26  & 6.92  & 0.8219  & 2.48  & 6.85  \\
          & DAB (auto-N-best+TS) & 0.8082  & 2.27  & 6.82  & 0.8140  & 2.23  & 6.81  & 0.8476  & 2.47  & 6.93  \bigstrut[b]\\
\cline{2-11}          & DAB (soft-N-best) & 0.6179  & 1.90  & 0.20  & 0.6140  & 1.88  & -0.29  & 0.6625  & 2.00  & 1.13  \bigstrut[t]\\
          & DAB (soft-N-best+GT) & 0.7792  & 2.27  & 6.10  & 0.7858  & 2.24  & 6.18  & 0.8094  & 2.43  & 6.06  \\
          & DAB (soft-N-best+TS) & 0.8045  & 2.26  & 6.35  & 0.8090  & 2.22  & 6.28  & 0.8428  & 2.46  & 6.54  \bigstrut[b]\\
\cline{2-11}          & DAB (learning-N-best) & 0.6187  & 1.90  & 0.31  & 0.6154  & 1.87  & -0.24  & 0.6562  & 1.98  & 0.93  \bigstrut[t]\\
          & DAB (learning-N-best+GT) & 0.7768  & 2.27  & 6.40  & 0.7799  & 2.24  & 6.30  & 0.8096  & 2.46  & 6.35  \\
          & DAB (learning-N-best+TS) & 0.8049  & 2.24  & 5.98  & 0.8100  & 2.20  & 5.90  & 0.8394  & 2.45  & 5.98  \bigstrut[b]\\
    \hline
    \multicolumn{1}{|r|}{\multirow{18}[12]{*}{20 dB}} & Noisy & 0.6622  & 2.03  & 3.81  & 0.6653  & 2.01  & 3.73  & 0.6860  & 2.12  & 3.71  \bigstrut[t]\\
          & DB    & 0.7539  & 2.09  & 4.90  & 0.7619  & 2.09  & 5.23  & 0.7792  & 2.21  & 5.29  \\
          & DAB (1-best) & 0.7768  & 2.25  & 7.25  & 0.7790  & 2.25  & 7.25  & 0.7967  & 2.31  & 7.13  \bigstrut[b]\\
\cline{2-11}          & DAB (all-channels) & 0.6196  & 1.88  & -0.31  & 0.6212  & 1.89  & -0.19  & 0.6213  & 1.89  & -0.50  \bigstrut[t]\\
          & DAB (all-channels+GT) & 0.7784  & 2.33  & 5.74  & 0.7834  & 2.32  & 5.68  & 0.7937  & 2.43  & 5.44  \\
          & DAB (all-channels+TS) & 0.8057  & 2.27  & 5.21  & 0.8113  & 2.25  & 5.02  & 0.8161  & 2.38  & 4.72  \bigstrut[b]\\
\cline{2-11}          & DAB (fixed-N-best) & 0.6583  & 1.96  & 1.03  & 0.6487  & 1.95  & 0.72  & 0.6553  & 1.94  & 0.57  \bigstrut[t]\\
          & DAB (fixed-N-best+GT) & 0.8011  & 2.35  & 6.60  & 0.8046  & 2.34  & 6.33  & 0.8183  & 2.42  & 6.65  \\
          & DAB (fixed-N-best+TS) & 0.8352  & 2.37  & 7.36  & 0.8346  & 2.34  & 7.08  & 0.8551  & 2.45  & 7.47  \bigstrut[b]\\
\cline{2-11}          & DAB (auto-N-best) & 0.6632  & 1.99  & 1.78  & 0.6504  & 1.97  & 1.20  & 0.6816  & 2.03  & 2.19  \bigstrut[t]\\
          & DAB (auto-N-best+GT) & 0.8098  & 2.40  & 7.28  & 0.8134  & 2.39  & 7.10  & 0.8257  & 2.48  & 7.21  \\
          & DAB (auto-N-best+TS) & 0.8361  & 2.39  & 7.67  & 0.8383  & 2.36  & 7.21  & 0.8515  & 2.48  & 7.35  \bigstrut[b]\\
\cline{2-11}          & DAB (soft-N-best) & 0.6610  & 1.99  & 1.52  & 0.6479  & 1.97  & 0.95  & 0.6810  & 2.03  & 2.00  \bigstrut[t]\\
          & DAB (soft-N-best+GT) & 0.8018  & 2.37  & 6.70  & 0.8034  & 2.36  & 6.30  & 0.8164  & 2.45  & 6.59  \\
          & DAB (soft-N-best+TS) & 0.8317  & 2.38  & 7.16  & 0.8338  & 2.35  & 6.74  & 0.8473  & 2.46  & 7.01  \bigstrut[b]\\
\cline{2-11}          & DAB (learning-N-best) & 0.6564  & 1.97  & 1.30  & 0.6491  & 1.96  & 0.93  & 0.6733  & 2.00  & 1.66  \bigstrut[t]\\
          & DAB (learning-N-best+GT) & 0.7968  & 2.38  & 6.74  & 0.8006  & 2.37  & 6.53  & 0.8139  & 2.47  & 6.57  \\
          & DAB (learning-N-best+TS) & 0.8309  & 2.36  & 6.74  & 0.8338  & 2.33  & 6.37  & 0.8447  & 2.47  & 6.43  \bigstrut[b]\\
    \hline
    \end{tabular}%
    }
  \label{tab:TS2}%
\end{table}%

\begin{table}[!t]
  \centering
  \caption{Results with 16 microphones per array in point source noise environments.}
  \scalebox{0.5}{
        \begin{tabular}{|r|l|rrr|rrr|rrr|}
    \hline
    \multicolumn{1}{|r|}{\multirow{2}[4]{*}{SNRatO}} & \multirow{2}[4]{*}{Comparison methods} & \multicolumn{3}{c|}{Babble} & \multicolumn{3}{c|}{Factory} & \multicolumn{3}{c|}{Volvo} \bigstrut\\
\cline{3-11}          &       & \multicolumn{1}{c}{STOI} & \multicolumn{1}{c}{PESQ} & \multicolumn{1}{c|}{SDR} & \multicolumn{1}{c}{STOI} & \multicolumn{1}{c}{PESQ} & \multicolumn{1}{c|}{SDR} & \multicolumn{1}{c}{STOI} & \multicolumn{1}{c}{PESQ} & \multicolumn{1}{c|}{SDR} \bigstrut\\
    \hline
    \multicolumn{1}{|r|}{\multirow{18}[12]{*}{-5 dB}} & Noisy & 0.4465  & 1.29  & -6.75  & 0.4336  & 1.19  & -6.08  & 0.6286  & 1.90  & -0.20  \bigstrut[t]\\
          & DB    & 0.5429  & 1.63  & -3.50  & 0.5250  & 1.51  & -2.22  & 0.7406  & 2.04  & 3.82  \\
          & DAB (1-best) & 0.5741  & 1.73  & -1.96  & 0.5512  & 1.59  & -1.52  & 0.7647  & 2.25  & 5.16  \bigstrut[b]\\
\cline{2-11}          & DAB (all-channels) & 0.4246  & 1.95  & -8.97  & 0.4194  & 1.73  & -8.10  & 0.5106  & 1.71  & -3.70  \bigstrut[t]\\
          & DAB (all-channels+GT) & 0.5756  & 1.70  & -2.41  & 0.5487  & 1.50  & -2.07  & 0.7424  & 2.22  & 3.95  \\
          & DAB (all-channels+TS) & 0.5954  & 1.70  & -2.43  & 0.5488  & 1.50  & -2.20  & 0.7775  & 2.22  & 3.30  \bigstrut[b]\\
\cline{2-11}          & DAB (fixed-N-best) & 0.4665  & 1.82  & -6.69  & 0.4619  & 1.66  & -5.83  & 0.5745  & 1.79  & -1.85  \bigstrut[t]\\
          & DAB (fixed-N-best+GT) & 0.5891  & 1.74  & -1.99  & 0.5619  & 1.58  & -1.68  & 0.7736  & 2.30  & 5.16  \\
          & DAB (fixed-N-best+TS) & 0.6065  & 1.74  & -1.65  & 0.5692  & 1.58  & -1.25  & 0.8124  & 2.33  & 5.57  \bigstrut[b]\\
\cline{2-11}          & DAB (auto-N-best) & 0.4753  & 1.93  & -6.60  & 0.4547  & 1.75  & -6.48  & 0.5773  & 1.86  & -1.18  \bigstrut[t]\\
          & DAB (auto-N-best+GT) & 0.6029  & 1.76  & -1.26  & 0.5707  & 1.55  & -1.21  & 0.7745  & 2.32  & 5.55  \\
          & DAB (auto-N-best+TS) & 0.6160  & 1.75  & -1.24  & 0.5696  & 1.55  & -1.23  & 0.8047  & 2.32  & 5.21  \bigstrut[b]\\
\cline{2-11}          & DAB (soft-N-best) & 0.4806  & 1.95  & -6.26  & 0.4601  & 1.76  & -6.08  & 0.5822  & 1.87  & -1.05  \bigstrut[t]\\
          & DAB (soft-N-best+GT) & 0.6035  & 1.77  & -1.15  & 0.5725  & 1.57  & -1.05  & 0.7681  & 2.29  & 5.08  \\
          & DAB (soft-N-best+TS) & 0.6164  & 1.76  & -1.15  & 0.5719  & 1.58  & -1.10  & 0.8013  & 2.32  & 4.92  \bigstrut[b]\\
\cline{2-11}          & DAB (learning-N-best) & 0.4606  & 1.92  & -7.34  & 0.4465  & 1.75  & -6.94  & 0.5654  & 1.83  & -1.80  \bigstrut[t]\\
          & DAB (learning-N-best+GT) & 0.5915  & 1.73  & -1.73  & 0.5621  & 1.53  & -1.54  & 0.7617  & 2.29  & 4.91  \\
          & DAB (learning-N-best+TS) & 0.6086  & 1.72  & -1.74  & 0.5604  & 1.53  & -1.72  & 0.7967  & 2.29  & 4.39  \bigstrut[b]\\
    \hline
    \multicolumn{1}{|r|}{\multirow{18}[12]{*}{5 dB}} & Noisy & 0.5678  & 1.68  & 0.11  & 0.5607  & 1.61  & 0.50  & 0.6550  & 1.99  & 2.69  \bigstrut[t]\\
          & DB    & 0.6975  & 1.87  & 2.80  & 0.6856  & 1.85  & 3.19  & 0.7695  & 2.14  & 4.78  \\
          & DAB (1-best) & 0.7232  & 2.05  & 5.22  & 0.7187  & 2.00  & 5.53  & 0.7939  & 2.31  & 7.42  \bigstrut[b]\\
\cline{2-11}          & DAB (all-channels) & 0.4942  & 1.79  & -3.59  & 0.4806  & 1.74  & -3.62  & 0.5207  & 1.74  & -2.69  \bigstrut[t]\\
          & DAB (all-channels+GT) & 0.7263  & 2.12  & 4.71  & 0.7168  & 2.06  & 4.98  & 0.7770  & 2.37  & 5.71  \\
          & DAB (all-channels+TS) & 0.7602  & 2.11  & 4.22  & 0.7481  & 2.05  & 4.31  & 0.8075  & 2.33  & 4.79  \bigstrut[b]\\
\cline{2-11}          & DAB (fixed-N-best) & 0.5522  & 1.80  & -1.80  & 0.5421  & 1.74  & -1.82  & 0.5889  & 1.83  & -0.92  \bigstrut[t]\\
          & DAB (fixed-N-best+GT) & 0.7473  & 2.14  & 5.28  & 0.7425  & 2.09  & 5.55  & 0.8117  & 2.42  & 7.22  \\
          & DAB (fixed-N-best+TS) & 0.7768  & 2.15  & 5.61  & 0.7734  & 2.11  & 5.81  & 0.8492  & 2.44  & 7.60  \bigstrut[b]\\
\cline{2-11}          & DAB (auto-N-best) & 0.5820  & 1.89  & -0.36  & 0.5901  & 1.86  & 0.39  & 0.6118  & 1.93  & 0.55  \bigstrut[t]\\
          & DAB (auto-N-best+GT) & 0.7601  & 2.17  & 5.94  & 0.7568  & 2.12  & 6.29  & 0.8187  & 2.46  & 7.68  \\
          & DAB (auto-N-best+TS) & 0.7835  & 2.17  & 6.02  & 0.7751  & 2.12  & 6.32  & 0.8455  & 2.45  & 7.54  \bigstrut[b]\\
\cline{2-11}          & DAB (soft-N-best) & 0.5834  & 1.90  & -0.47  & 0.5915  & 1.85  & 0.25  & 0.6122  & 1.93  & 0.42  \bigstrut[t]\\
          & DAB (soft-N-best+GT) & 0.7534  & 2.16  & 5.48  & 0.7510  & 2.10  & 5.83  & 0.8081  & 2.42  & 6.85  \\
          & DAB (soft-N-best+TS) & 0.7797  & 2.16  & 5.61  & 0.7714  & 2.11  & 5.91  & 0.8420  & 2.44  & 7.09  \bigstrut[b]\\
\cline{2-11}          & DAB (learning-N-best) & 0.5605  & 1.86  & -1.27  & 0.5617  & 1.82  & -0.86  & 0.5975  & 1.90  & -0.09  \bigstrut[t]\\
          & DAB (learning-N-best+GT) & 0.7462  & 2.15  & 5.37  & 0.7397  & 2.09  & 5.71  & 0.8034  & 2.43  & 6.95  \\
          & DAB (learning-N-best+TS) & 0.7806  & 2.16  & 5.40  & 0.7672  & 2.10  & 5.38  & 0.8378  & 2.43  & 6.57  \bigstrut[b]\\
    \hline
    \multicolumn{1}{|r|}{\multirow{18}[12]{*}{15 dB}} & Noisy & 0.6394  & 1.92  & 2.71  & 0.6405  & 1.90  & 2.76  & 0.6700  & 2.02  & 3.16  \bigstrut[t]\\
          & DB    & 0.7534  & 2.11  & 4.85  & 0.7596  & 2.10  & 5.01  & 0.7767  & 2.21  & 5.21  \\
          & DAB (1-best) & 0.7868  & 2.26  & 7.48  & 0.7886  & 2.23  & 7.39  & 0.8024  & 2.32  & 7.63  \bigstrut[b]\\
\cline{2-11}          & DAB (all-channels) & 0.5215  & 1.76  & -2.52  & 0.5152  & 1.73  & -2.68  & 0.5183  & 1.76  & -2.69  \bigstrut[t]\\
          & DAB (all-channels+GT) & 0.7770  & 2.36  & 6.16  & 0.7763  & 2.33  & 6.12  & 0.7871  & 2.43  & 5.98  \\
          & DAB (all-channels+TS) & 0.8173  & 2.35  & 5.77  & 0.8189  & 2.31  & 5.66  & 0.8173  & 2.41  & 5.14  \bigstrut[b]\\
\cline{2-11}          & DAB (fixed-N-best) & 0.5924  & 1.82  & -0.69  & 0.5886  & 1.80  & -0.78  & 0.5911  & 1.83  & -0.81  \bigstrut[t]\\
          & DAB (fixed-N-best+GT) & 0.8015  & 2.35  & 6.81  & 0.7999  & 2.32  & 6.77  & 0.8177  & 2.44  & 7.23  \\
          & DAB (fixed-N-best+TS) & 0.8434  & 2.40  & 7.68  & 0.8419  & 2.35  & 7.54  & 0.8591  & 2.48  & 7.87  \bigstrut[b]\\
\cline{2-11}          & DAB (auto-N-best) & 0.6503  & 2.01  & 2.22  & 0.6042  & 1.89  & 0.39  & 0.6602  & 2.03  & 2.39  \bigstrut[t]\\
          & DAB (auto-N-best+GT) & 0.8156  & 2.40  & 7.93  & 0.8088  & 2.38  & 7.48  & 0.8292  & 2.46  & 7.94  \\
          & DAB (auto-N-best+TS) & 0.8405  & 2.41  & 8.08  & 0.8422  & 2.37  & 7.50  & 0.8502  & 2.47  & 8.12  \bigstrut[b]\\
\cline{2-11}          & DAB (soft-N-best) & 0.6499  & 2.01  & 2.13  & 0.6042  & 1.90  & 0.30  & 0.6595  & 2.03  & 2.27  \bigstrut[t]\\
          & DAB (soft-N-best+GT) & 0.8088  & 2.39  & 7.44  & 0.8009  & 2.35  & 6.84  & 0.8226  & 2.44  & 7.45  \\
          & DAB (soft-N-best+TS) & 0.8379  & 2.40  & 7.79  & 0.8385  & 2.37  & 7.09  & 0.8477  & 2.46  & 7.85  \bigstrut[b]\\
\cline{2-11}          & DAB (learning-N-best) & 0.6272  & 1.95  & 1.25  & 0.5862  & 1.85  & -0.33  & 0.6340  & 1.98  & 1.18  \bigstrut[t]\\
          & DAB (learning-N-best+GT) & 0.8028  & 2.39  & 7.30  & 0.7966  & 2.36  & 6.97  & 0.8148  & 2.45  & 7.28  \\
          & DAB (learning-N-best+TS) & 0.8415  & 2.42  & 7.28  & 0.8392  & 2.37  & 6.89  & 0.8500  & 2.49  & 7.22  \bigstrut[b]\\
    \hline
    \end{tabular}%
  \label{tab:TS4}%
  }
\end{table}%

We list the performance of the comparison methods in the diffuse noise and point source noise environments in Tables \ref{tab:TS2} and \ref{tab:TS4} respectively. From the tables, we see that the DAB variants given the TS module or the ground-truth time delay outperform the DB baseline significantly in terms of all evaluation metrics. Even the simplest ``DAB+1-best'' is better than the DB baseline, which demonstrates the advantage of the ad-hoc microphone array.

\begin{figure*}[!t]
\centering
\includegraphics[width=17cm]{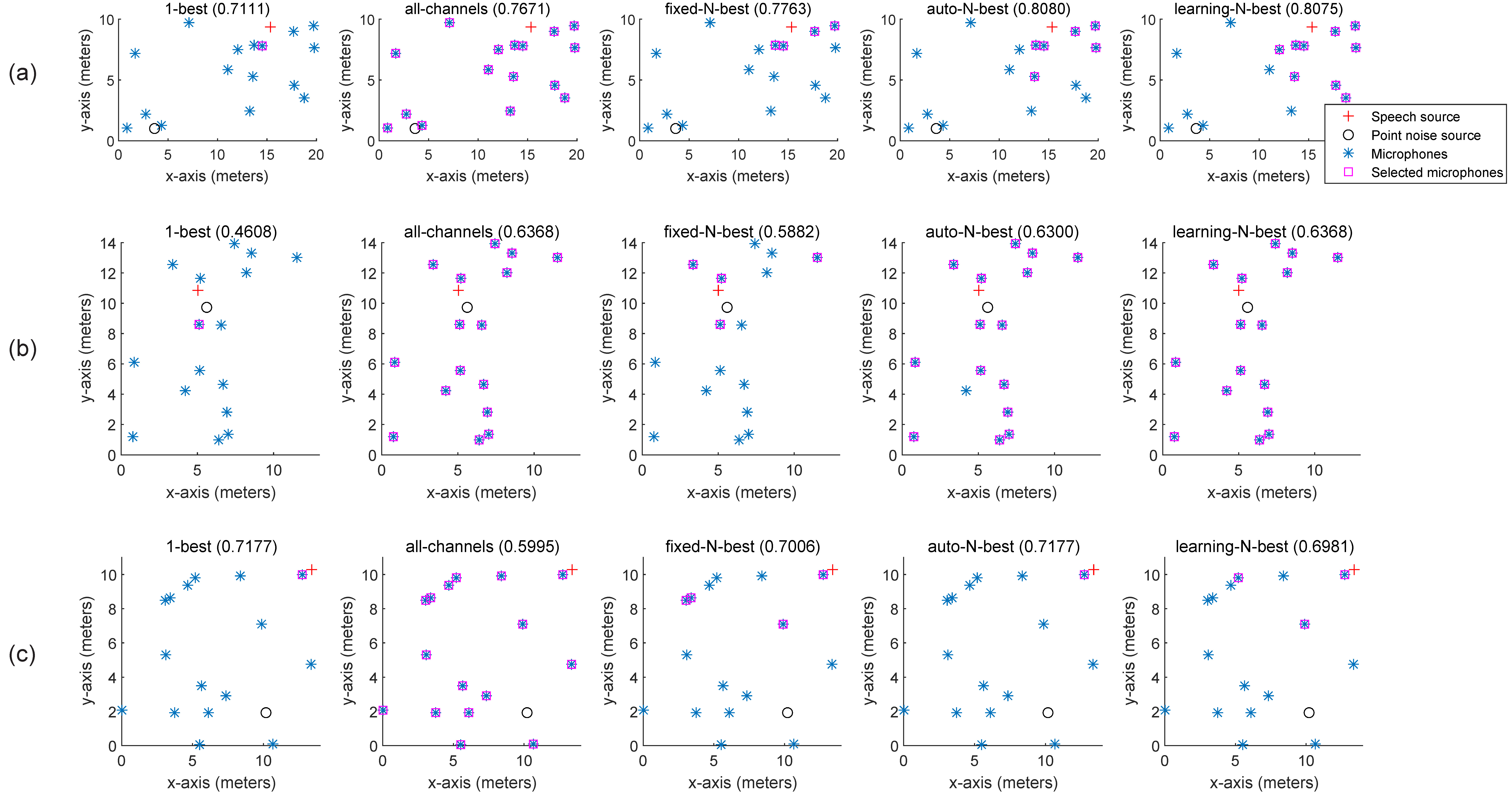}
\caption{Examples of channel-selection results in the babble point source noise environments at the SNRatO of $-5$ dB, where the number in the brackets of the title of each sub-figure is the STOI value. (a) Room size: $20\times 10\times 3.5\mbox{ m}^3$. (b) Room size: $13\times 14\times 3.1 \mbox{ m}^3$. (c) Room size: $14\times 11\times 3.1 \mbox{ m}^3$} \label{fig:channel_s}
\end{figure*}

We compare the DB variants with the TS module or the ground-truth time delay for studying the effectiveness of the channel selection algorithms. We find that ``auto-\textit{N}-best'' performs the best among the channel selection algorithms in most cases, followed by ``soft-\textit{N}-best''.
 The ``learning-\textit{N}-best'' and ``fixed-\textit{N}-best'' algorithms perform equivalently well in general, both of which perform better than the ``all-channels'' algorithm. Although ``1-best'' performs the poorest in terms of STOI and PESQ, it usually produces good SDR scores that are comparable to those produced by ``auto-\textit{N}-best''. Note that although ``learning-\textit{N}-best'' seems an advanced technique, this advantage does not transfer to superior performance. This may be caused by \eqref{eq:leanring} which is an expansion of the channel selection result of ``auto-\textit{N}-best''. This problem needs further investigation in the future.
 Comparing ``auto-\textit{N}-best'' and ``soft-\textit{N}-best'', we further find that different amplitude ranges of the channels affect the performance, though this phenomenon is not so obvious due to that the nonzero weights do not vary in a large range. As a byproduct, the idea of ``soft-\textit{N}-best'' is a way of synchronizing the adaptive gain controllers of the devices. The synchronization of the adaptive gain controllers is not the focus of this paper, hence we leave it for the future study.

We compare the ``DAB+channel selection method'', ``DAB+channel selection method+GT'', and ``DAB+channel selection method+TS'' given different channel selection methods for studying the effectiveness of the time synchronization module. We find that the DAB without the TS module does not work at all when there exists a serious time unsynchronization problem caused by devices. ``DAB+channel selection method+TS'' performs better than ``DAB+channel selection method+GT'' in terms of STOI, and is equivalently good with the latter in terms of PESQ and SDR in all SNRatO levels, even though the latter was given the ground-truth time delay caused by devices. This phenomenon demonstrates the effectiveness of the proposed TS module. It also implies that the time unsynchronization problem caused by different locations of the microphones affects the performance, though not so serious.

Figure \ref{fig:channel_s} shows three examples of the channel selection results, of which we find some interesting phenomena after looking into the details. Figure \ref{fig:channel_s}a is a typical scenario where the speech source is far away from the noise point source. We see clearly from the figure that, although ``1-best'' is relative much poorer than the other algorithms, all comparison algorithms perform not so bad according to the absolute STOI scores, since that the SNRs at many selectted microphones are relative high. Figure \ref{fig:channel_s}b is a special scenario where the speech source is very close to the noise point source. Therefore, the SNRs at all microphones are low. As shown in the figure, it is better to select most microphones as ``auto-{\it{N}}-best'' and ``learning-{\it{N}}-best'' do in this case, otherwise the performance is rather poor as ``1-best'' yields. Figure \ref{fig:channel_s}c is a special scenario where there is a microphone very close to the speech source. It can be seen from the channel selection result that the best way is to select the closest microphone, while ``all-channels'' perform much poorer than the other channel selection algorithms. To summarize the above phenomena, we see that the adaptive channel selection algorithms, i.e. ``auto-{\it{N}}-best'' and ``learning-{\it{N}}-best'', always produce top performance among the comparison algorithms.

\subsection{Effect of the number of the microphones in an array}\label{subsec:number_phone}

To study how the number of the microphones in an array affects the performance, we repeated the experimental setting in Section \ref{subsec:exp_set} except that the number of the microphones in an array was reduced to 4. Because the experimental phenomena were consistent across different SNRatO levels and noise types, we list the comparison results of only one test scenario in Tables \ref{tab:addlabel_4ch_di} and \ref{tab:addlabel_4ch_point} for saving the space of the paper. From the tables, we see that, even if the number of the microphones in an array was limited, the DAB variants still perform equivalently well with DB except ``DAB+1-best''. Comparing Tables \ref{tab:addlabel_4ch_di} and \ref{tab:addlabel_4ch_point} with Tables \ref{tab:TS2} and \ref{tab:TS4}, we also find that DAB benefits much more than DB from the increase of the number of the microphones. We take the results in the diffuse noise environment as an example. The STOI score of ``DAB+auto-\textit{N}-best+TS'' is improved by relatively 20.22\% when the number of the microphones is increased from 4 to 16, while the relative improvement about DB is only 2.56\%.

\begin{table}[!t]
  \centering
  \caption{Results with 4 microphones per array in babble diffuse noise environment at an SNRatO of 10 dB.}
  \scalebox{0.6}{
    \begin{tabular}{|r|l|rrr|}
    \hline
    \multicolumn{1}{|r|}{\multirow{2}[4]{*}{SNRatO}} & \multirow{2}[4]{*}{Comparison methods} & \multicolumn{3}{c|}{Babble} \bigstrut\\
\cline{3-5}          &       & \multicolumn{1}{c}{STOI} & \multicolumn{1}{c}{PESQ} & \multicolumn{1}{c|}{SDR} \bigstrut\\
    \hline
    \multicolumn{1}{|r|}{\multirow{8}[2]{*}{10 dB}} & Noisy & 0.5919 &	1.80 &	0.99  \bigstrut[t]\\
          & DB    & 0.6830  & 1.91  & 3.14  \\
          & DAB (1-best) & 0.6400  & 1.86  & 2.38  \\
          & DAB (all-channels+TS) & 0.7154  & 1.92  & 2.70  \\
          & DAB (fixed-N-best+TS) & 0.6821  & 1.90  & 2.58  \\
          & DAB (auto-N-best+TS) & 0.7112  & 1.93  & 3.01  \\
          & DAB (soft-N-best+TS) & 0.7013  & 1.91  & 2.27  \\
          & DAB (learning-N-best+TS) & 0.7135  & 1.92  & 2.82  \bigstrut[b]\\
    \hline
    \end{tabular}%
    }
  \label{tab:addlabel_4ch_di}%
\end{table}%

\begin{table}[!t]
  \centering
  \caption{Results with 4 microphones per array in babble point source noise environment at an SNRatO of $-5$ dB.}
  \scalebox{0.6}{
    \begin{tabular}{|r|l|rrr|}
    \hline
    \multicolumn{1}{|r|}{\multirow{2}[4]{*}{SNRatO}} & \multirow{2}[4]{*}{Comparison methods} & \multicolumn{3}{c|}{Babble} \bigstrut\\
\cline{3-5}          &       & \multicolumn{1}{c}{STOI} & \multicolumn{1}{c}{PESQ} & \multicolumn{1}{c|}{SDR} \bigstrut\\
    \hline
    \multicolumn{1}{|r|}{\multirow{8}[2]{*}{-5 dB}} & Noisy & 0.4576  & 1.56  & -6.14  \bigstrut[t]\\
          & DB    & 0.5079  & 1.47  & -5.70  \\
          & DAB (1-best) & 0.4996  & 1.57  & -5.10  \\
          & DAB (all-channels+TS) & 0.5056  & 1.49  & -6.80  \\
          & DAB (fixed-N-best+TS) & 0.5068  & 1.55  & -5.72  \\
          & DAB (auto-N-best+TS) & 0.5111  & 1.51  & -6.28  \\
          & DAB (soft-N-best+TS) & 0.5140  & 1.53  & -6.19  \\
          & DAB (learning-N-best+TS) & 0.5064  & 1.50  & -6.67  \bigstrut[b]\\
    \hline
    \end{tabular}%
    }
  \label{tab:addlabel_4ch_point}%
\end{table}%

\subsection{Effect of hyperparameter $\gamma$}\label{subsec2:hyper}

\begin{figure}[!t]
\centering
\includegraphics[width=9cm]{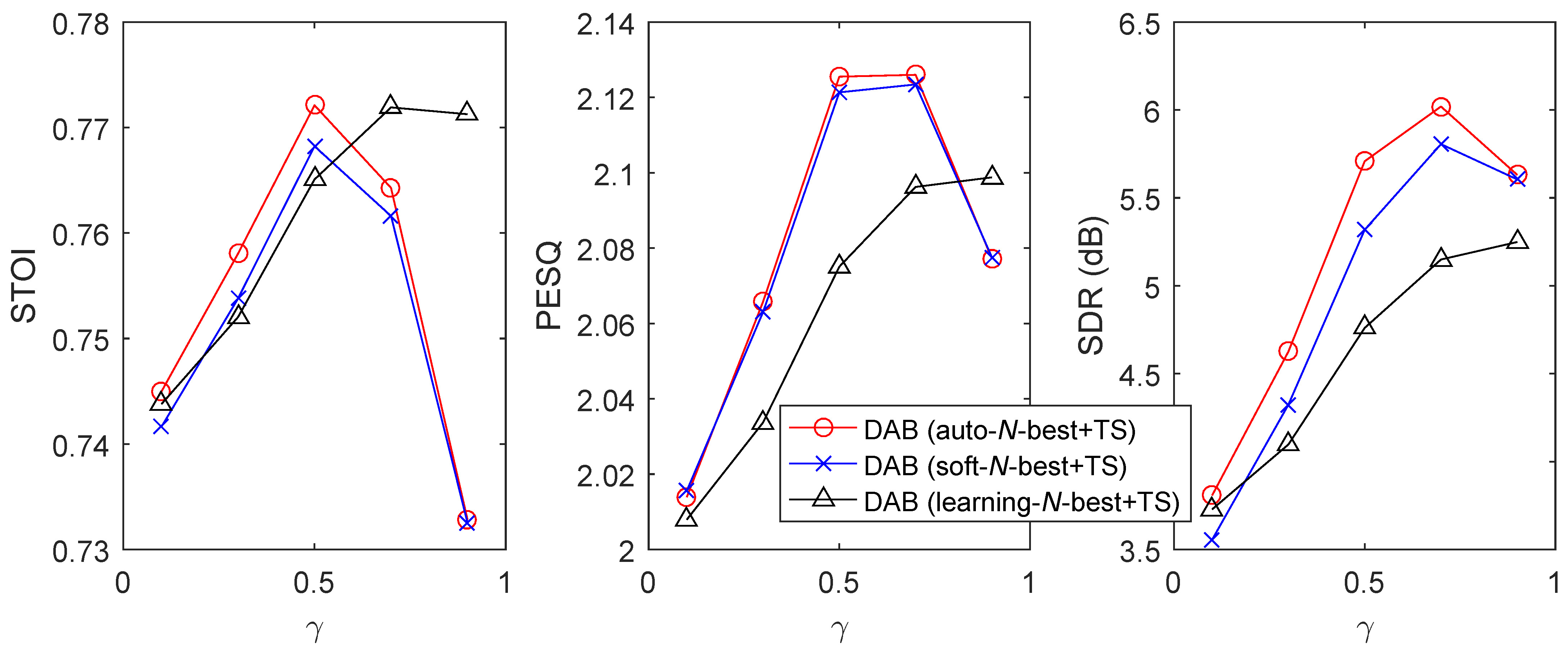}
\caption{Effect of hyperparameter $\gamma$ in the babble diffuse noise environment at the SNRatO of $10$ dB.} \label{fig:gamma_reverb}
\end{figure}

\begin{figure}[!t]
\centering
\includegraphics[width=9cm]{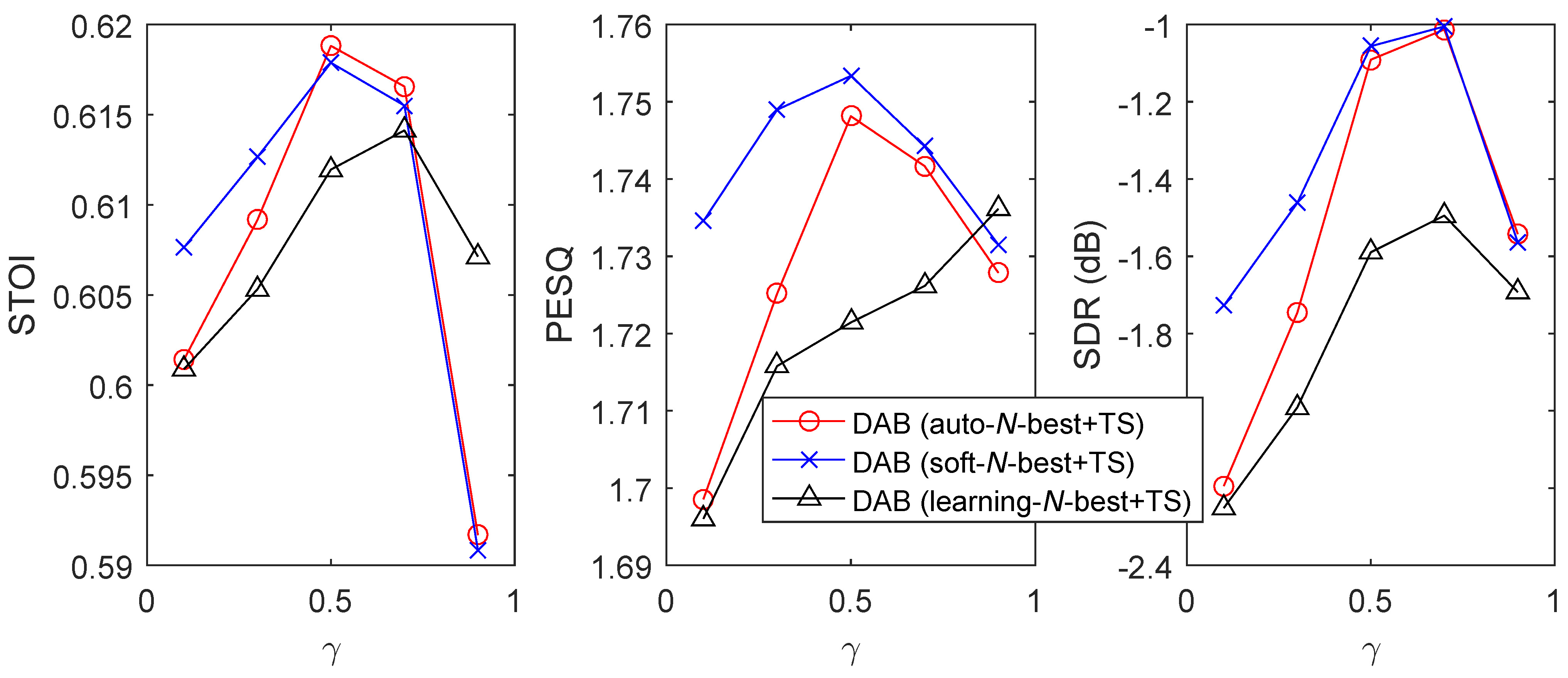}
\caption{Effect of hyperparameter $\gamma$ in the babble point source noise environment at the SNRatO of $-5$ dB.} \label{fig:gamma}
\end{figure}

To study how the hyperparameter $\gamma$ affects the performance of ``DAB+auto-\textit{N}-best+TS'', ``DAB+soft-\textit{N}-best+TS'', and ``DAB+learning-\textit{N}-best+TS'', we tune $\gamma$ from $\{0.1, 0.3, 0.5, 0.7, 0.9\}$. To save the space of the paper, we only show the results in the babble noise environments at the lowest SNRatO levels in Figs. \ref{fig:gamma_reverb} and \ref{fig:gamma}. From the figures, we observe that
 ``DAB+auto-\textit{N}-best+TS'' and ``DAB+soft-\textit{N}-best+TS'' perform similarly if $\gamma$ is well-tuned, both of which are better than  ``DAB+soft-\textit{N}-best+TS''. The working range of $\gamma$ is $[0.5, 0.7]$ for ``DAB+auto-\textit{N}-best+TS'' and ``DAB+soft-\textit{N}-best+TS'', and $[0.7, 0.9]$ for ``DAB+learning-\textit{N}-best+TS''.

\subsection{Effect of handcrafted features on performance}

All above experiments were conducted with the eSTFT feature. To study how the handcrafted features affect the performance, we compared the DAB models that took eSTFT and MRCG respectively as the input of the channel-selection model $g(\cdot)$ in the babble noise environments at the lowest SNRatO levels. From the comparison results in Tables \ref{fig:STFT} and \ref{fig:MRCG}, we see that eSTFT is slightly better than MRCG in the diffuse noise environment, and significantly outperforms MRCG in the point source noise environment. The effect of different acoustic features for ``DAB+1-best+TS'' in the point source noise environment is remarkable, which manifests the importance of designing a good handcrafted feature. We also observe that the advantage of the adaptive channel selection algorithms over ``DAB+1-best+TS'' and ``DAB+fixed-\textit{N}-best+TS'' is consistent across the two acoustic features, which demonstrates the robustness of the proposed channel selection algorithms to the choice of the acoustic features.

\begin{table}[!t]
  \centering
  \caption{Comparison results of handcrafted features for the DAB variants with 16 microphones per array in babble diffuse noise environment at an SNRatO of $10$ dB.}
  \scalebox{0.6}{
    \begin{tabular}{|c|l|rrr|}
    \hline
    \multirow{2}[4]{*}{SNRatO} & \multicolumn{1}{c|}{\multirow{2}[4]{*}{Comparison methods}} & \multicolumn{3}{c|}{Babble} \bigstrut\\
\cline{3-5}          &       & \multicolumn{1}{c}{STOI} & \multicolumn{1}{c}{PESQ} & \multicolumn{1}{c|}{SDR} \bigstrut\\
    \hline
    \multirow{10}[2]{*}{10 dB} & 1-best (eSTFT) & 0.7154  & 2.06  & 5.14  \bigstrut[t]\\
          & 1-best (MRCG) & 0.7101  & 2.05  & 5.01  \\
          & fixed-N-best+TS  (eSTFT) & 0.7675  & 2.11  & 5.18  \\
          & fixed-N-best+TS  (MRCG) & 0.7617  & 2.10  & 4.96  \\
          & auto-N-best+TS (eSTFT) & 0.7696  & 2.12  & 5.45  \\
          & auto-N-best+TS (MRCG) & 0.7654  & 2.11  & 5.28  \\
          & soft-N-best+TS (eSTFT) & 0.7659  & 2.12  & 5.11  \\
          & soft-N-best+TS (MRCG) & 0.7621  & 2.11  & 4.88  \\
          & learning-N-best+TS (eSTFT) & 0.7631  & 2.07  & 4.55  \\
          & learning-N-best+TS (MRCG) & 0.7606  & 2.07  & 4.48  \bigstrut[b]\\
    \hline
    \end{tabular}%
    }
  \label{fig:STFT}%
\end{table}%

\begin{table}[!t]
  \centering
  \caption{Comparison results of handcrafted features for the DAB variants with 16 microphones per array in babble point source noise environment at an SNRatO of $-5$ dB.}
  \scalebox{0.6}{
    \begin{tabular}{|c|l|rrr|}
    \hline
    \multirow{2}[4]{*}{SNRatO} & \multicolumn{1}{c|}{\multirow{2}[4]{*}{Comparison methods}} & \multicolumn{3}{c|}{Babble} \bigstrut\\
\cline{3-5}          &       & \multicolumn{1}{c}{STOI} & \multicolumn{1}{c}{PESQ} & \multicolumn{1}{c|}{SDR} \bigstrut\\
    \hline
    \multirow{10}[2]{*}{-5 dB} & 1-best (eSTFT) & 0.5741  & 1.73  & -1.96  \bigstrut[t]\\
          & 1-best (MRCG) & 0.5267  & 1.67  & -3.77  \\
          & fixed-N-best+TS  (eSTFT) & 0.6065  & 1.74  & -1.65  \\
          & fixed-N-best+TS  (MRCG) & 0.5838  & 1.67  & -2.55  \\
          & auto-N-best+TS (eSTFT) & 0.6160  & 1.75  & -1.24  \\
          & auto-N-best+TS (MRCG) & 0.5918  & 1.70  & -2.14  \\
          & soft-N-best+TS (eSTFT) & 0.6164  & 1.76  & -1.15  \\
          & soft-N-best+TS (MRCG) & 0.5934  & 1.70  & -2.10  \\
          & learning-N-best+TS (eSTFT) & 0.6086  & 1.72  & -1.74  \\
          & learning-N-best+TS (MRCG) & 0.5956  & 1.70  & -2.27  \bigstrut[b]\\
    \hline
    \end{tabular}%
    }
  \label{fig:MRCG}%
\end{table}%

\section{Conclusions and future work}\label{sec:conclusion}

In this paper, we have proposed deep ad-hoc beamforming, which is to our knowledge the first deep learning method designed for ad-hoc microphone arrays.\footnote{This claim was made according to the fact that the core idea of the paper has been put on arXiv \citep{zhang2018deep} in January 2019.} DAB has the following novel aspects. First, DAB employs an ad-hoc microphone array to pick up speech signals, which has a potential to enhance the speech signals with equally high quality in a range where the array covers. It may also significantly improve the SNR at the microphone receivers by physically placing some microphones close to the speech source in probability. Second, DAB employs a channel-selection algorithm to reweight the estimated speech signals with a sparsity constraint, which groups a handful microphones around the speech source into a local microphone array. We have developed several channel-selection algorithms as well. Third, we have developed a time synchronization framework based on time delay estimators and the supervised 1-best channel selection. At last, we emphasized the importance of acoustic features to DAB by carrying out the first study on how different acoustic features affect the performance.

Besides the above novelties and advantages, the proposed DAB is flexible in incorporating new development of DNN-based single channel speech processing techniques, since that its model is trained in the single-channel fashion. Its test process is also flexible in incorporating any number of microphones without retraining or revising the model, which meets the requirement of real-world applications. Moreover, although we applied DAB to speech enhancement as an example, we may apply it to other tasks as well by replacing the deep beamforming to other task-specific algorithms.

 We have conducted extensive experiments in the scenario where the location of the speech source is far-field, random, and blind to the microphones. Experimental results in both the diffuse noise and point source noise environments demonstrate that DAB outperforms its MVDR based deep beamforming counterpart by a large margin given enough number of microphones. The conclusion is consistent across different acoustic features.

 The research on DAB is only at the beginning. There are too many open problems. Here we list some urgent topics as follows. (i) How to synchronize microphones when the clock rates and power amplifiers at different devices are different. (ii) How to design new spatial acoustic features for ad-hoc microphone arrays beyond interaural time difference or interaural level difference. (iii) How to design a model that can be trained with multichannel data collected from ad-hoc microphone arrays, and generalize well to fundamentally different ad-hoc microphone arrays in the test stage.
 (iv) How to handle a large number of microphones (e.g. over 100 microphones) for a large room that contains many complicated acoustic environments.


\section*{Acknowledgments}
The author would like to thank Prof. DeLiang Wang for helpful discussions.

This work was supported in part by the National Key Research and Development Program of China
        under Grant No. 2018AAA0102200, in part by National Science Foundation
        of China under Grant No. 61831019, 61671381, in part by the Project of the Science, Technology, and Innovation Commission of
Shenzhen Municipality under grant No. JCYJ20170815161820095, and in part by the Open Research Project of the State Key Laboratory of Media Convergence and Communication, Communication University of China, China under Grant No. SKLMCC2020KF009.

\appendix
\section{}
\begin{proof}
  We denote the energy of the direct sound and additive noise components of the test utterance at the $i$-th channel as $X_i$ and $N_i$ respectively, i.e. $X = \sum_t |x|_{\rm{time}}(t)$ and $N = \sum_t|n|_{\rm{time}}(t)$. Our core idea is to filter out the signals of the channels whose clean speech satisfies:
    \begin{equation}\label{eq:xxq}
X_i<\gamma X_{*}
\end{equation}
   Under the assumptions that the estimated weights are perfect and that the statistics of the noise components are consistent across the channels, we have
       \begin{eqnarray}\label{eq:xxq1}
&&q_i=\frac{S_i}{S_i+N_*},\quad q_{*}=\frac{S_{*}}{S_{*}+N_{*}}\label{eq:xxq2}
\end{eqnarray}
Substituting \eqref{eq:xxq1} into \eqref{eq:xxq} derives \eqref{14}.
\end{proof}

\section*{References}

\bibliographystyle{elsarticle-harv}
\bibliography{zxlrefs2,zxlrefs,mywork}

\begin{thebibliography}{68}
\expandafter\ifx\csname natexlab\endcsname\relax\def\natexlab#1{#1}\fi
\providecommand{\url}[1]{\texttt{#1}}
\providecommand{\href}[2]{#2}
\providecommand{\path}[1]{#1}
\providecommand{\DOIprefix}{doi:}
\providecommand{\ArXivprefix}{arXiv:}
\providecommand{\URLprefix}{URL: }
\providecommand{\Pubmedprefix}{pmid:}
\providecommand{\doi}[1]{\href{http://dx.doi.org/#1}{\path{#1}}}
\providecommand{\Pubmed}[1]{\href{pmid:#1}{\path{#1}}}
\providecommand{\bibinfo}[2]{#2}
\ifx\xfnm\relax \def\xfnm[#1]{\unskip,\space#1}\fi
\bibitem[{Bai et~al.(2020{\natexlab{a}})Bai, Zhang \& Chen}]{bai2020cosine}
\bibinfo{author}{Bai, Z.}, \bibinfo{author}{Zhang, X.-L.}, \&
  \bibinfo{author}{Chen, J.} (\bibinfo{year}{2020}{\natexlab{a}}).
\newblock \bibinfo{title}{Cosine metric learning based speaker verification}.
\newblock {\it \bibinfo{journal}{Speech Communication}\/},  {\it
  \bibinfo{volume}{118}\/}, \bibinfo{pages}{10--20}.
\bibitem[{Bai et~al.(2020{\natexlab{b}})Bai, Zhang \& Chen}]{bai2020partial}
\bibinfo{author}{Bai, Z.}, \bibinfo{author}{Zhang, X.-L.}, \&
  \bibinfo{author}{Chen, J.} (\bibinfo{year}{2020}{\natexlab{b}}).
\newblock \bibinfo{title}{Partial auc optimization based deep speaker
  embeddings with class-center learning for text-independent speaker
  verification}.
\newblock In {\it \bibinfo{booktitle}{ICASSP 2020-2020 IEEE International
  Conference on Acoustics, Speech and Signal Processing (ICASSP)}\/} (pp.
  \bibinfo{pages}{6819--6823}).
\newblock \bibinfo{organization}{IEEE}.
\bibitem[{Bai et~al.(2020{\natexlab{c}})Bai, Zhang \& Chen}]{bai2020speaker}
\bibinfo{author}{Bai, Z.}, \bibinfo{author}{Zhang, X.-L.}, \&
  \bibinfo{author}{Chen, J.} (\bibinfo{year}{2020}{\natexlab{c}}).
\newblock \bibinfo{title}{Speaker verification by partial auc optimization with
  mahalanobis distance metric learning}.
\newblock {\it \bibinfo{journal}{IEEE/ACM Transactions on Audio, Speech, and
  Language Processing}\/}, .
\bibitem[{Carter(1987)}]{carter1987coherence}
\bibinfo{author}{Carter, G.~C.} (\bibinfo{year}{1987}).
\newblock \bibinfo{title}{Coherence and time delay estimation}.
\newblock {\it \bibinfo{journal}{Proceedings of the IEEE}\/},  {\it
  \bibinfo{volume}{75}\/}, \bibinfo{pages}{236--255}.
\bibitem[{Chen et~al.(2006)Chen, Benesty \& Huang}]{chen2006time}
\bibinfo{author}{Chen, J.}, \bibinfo{author}{Benesty, J.}, \&
  \bibinfo{author}{Huang, Y.~A.} (\bibinfo{year}{2006}).
\newblock \bibinfo{title}{Time delay estimation in room acoustic environments:
  an overview}.
\newblock {\it \bibinfo{journal}{EURASIP Journal on Advances in Signal
  Processing}\/},  {\it \bibinfo{volume}{2006}\/}, \bibinfo{pages}{026503}.
\bibitem[{Chen et~al.(2014)Chen, Wang \& Wang}]{chen2014feature}
\bibinfo{author}{Chen, J.}, \bibinfo{author}{Wang, Y.}, \&
  \bibinfo{author}{Wang, D.~L.} (\bibinfo{year}{2014}).
\newblock \bibinfo{title}{A feature study for classification-based speech
  separation at very low signal-to-noise ratio}.
\newblock {\it \bibinfo{journal}{IEEE/ACM Trans. Audio, Speech, Lang.
  Process.}\/},  {\it \bibinfo{volume}{22}\/}, \bibinfo{pages}{1993--2002}.
\bibitem[{Delfarah \& Wang(2017)}]{delfarah2017features}
\bibinfo{author}{Delfarah, M.}, \& \bibinfo{author}{Wang, D.}
  (\bibinfo{year}{2017}).
\newblock \bibinfo{title}{Features for masking-based monaural speech separation
  in reverberant conditions}.
\newblock {\it \bibinfo{journal}{IEEE/ACM Transactions on Audio, Speech, and
  Language Processing}\/},  {\it \bibinfo{volume}{25}\/},
  \bibinfo{pages}{1085--1094}.
\bibitem[{Ditter \& Gerkmann(2020)}]{ditter2020multi}
\bibinfo{author}{Ditter, D.}, \& \bibinfo{author}{Gerkmann, T.}
  (\bibinfo{year}{2020}).
\newblock \bibinfo{title}{A multi-phase gammatone filterbank for speech
  separation via tasnet}.
\newblock In {\it \bibinfo{booktitle}{ICASSP 2020-2020 IEEE International
  Conference on Acoustics, Speech and Signal Processing (ICASSP)}\/} (pp.
  \bibinfo{pages}{36--40}).
\newblock \bibinfo{organization}{IEEE}.
\bibitem[{Du et~al.(2014)Du, Tu, Xu, Dai \& Lee}]{du2014speech}
\bibinfo{author}{Du, J.}, \bibinfo{author}{Tu, Y.}, \bibinfo{author}{Xu, Y.},
  \bibinfo{author}{Dai, L.}, \& \bibinfo{author}{Lee, C.-H.}
  (\bibinfo{year}{2014}).
\newblock \bibinfo{title}{Speech separation of a target speaker based on deep
  neural networks}.
\newblock In {\it \bibinfo{booktitle}{Proc. IEEE Int. Conf. Signal Process.}\/}
  (pp. \bibinfo{pages}{473--477}).
\bibitem[{Erdogan et~al.(2016)Erdogan, Hershey, Watanabe, Mandel \&
  Le~Roux}]{erdogan2016improved}
\bibinfo{author}{Erdogan, H.}, \bibinfo{author}{Hershey, J.~R.},
  \bibinfo{author}{Watanabe, S.}, \bibinfo{author}{Mandel, M.~I.}, \&
  \bibinfo{author}{Le~Roux, J.} (\bibinfo{year}{2016}).
\newblock \bibinfo{title}{Improved {MVDR} beamforming using single-channel mask
  prediction networks.}
\newblock In {\it \bibinfo{booktitle}{Interspeech}\/} (pp.
  \bibinfo{pages}{1981--1985}).
\bibitem[{Guariglia(2019)}]{guariglia2019primality}
\bibinfo{author}{Guariglia, E.} (\bibinfo{year}{2019}).
\newblock \bibinfo{title}{Primality, fractality, and image analysis}.
\newblock {\it \bibinfo{journal}{Entropy}\/},  {\it \bibinfo{volume}{21}\/},
  \bibinfo{pages}{304}.
\bibitem[{Guariglia \& Silvestrov(2016)}]{guariglia2016fractional}
\bibinfo{author}{Guariglia, E.}, \& \bibinfo{author}{Silvestrov, S.}
  (\bibinfo{year}{2016}).
\newblock \bibinfo{title}{{Fractional-wavelet analysis of positive definite
  distributions and wavelets on D'(C)}}.
\newblock In {\it \bibinfo{booktitle}{Engineering Mathematics II}\/} (pp.
  \bibinfo{pages}{337--353}).
\newblock \bibinfo{publisher}{Springer}.
\bibitem[{Guido(2018{\natexlab{a}})}]{guido2018fusing}
\bibinfo{author}{Guido, R.~C.} (\bibinfo{year}{2018}{\natexlab{a}}).
\newblock \bibinfo{title}{Fusing time, frequency and shape-related information:
  Introduction to the discrete shapelet transform’s second generation
  (dst-ii)}.
\newblock {\it \bibinfo{journal}{Information Fusion}\/},  {\it
  \bibinfo{volume}{41}\/}, \bibinfo{pages}{9--15}.
\bibitem[{Guido(2018{\natexlab{b}})}]{guido2018tutorial}
\bibinfo{author}{Guido, R.~C.} (\bibinfo{year}{2018}{\natexlab{b}}).
\newblock \bibinfo{title}{A tutorial review on entropy-based handcrafted
  feature extraction for information fusion}.
\newblock {\it \bibinfo{journal}{Information Fusion}\/},  {\it
  \bibinfo{volume}{41}\/}, \bibinfo{pages}{161--175}.
\bibitem[{Heusdens et~al.(2012)Heusdens, Zhang, Hendriks, Zeng \&
  Kleijn}]{heusdens2012distributed}
\bibinfo{author}{Heusdens, R.}, \bibinfo{author}{Zhang, G.},
  \bibinfo{author}{Hendriks, R.~C.}, \bibinfo{author}{Zeng, Y.}, \&
  \bibinfo{author}{Kleijn, W.~B.} (\bibinfo{year}{2012}).
\newblock \bibinfo{title}{Distributed {MVDR} beamforming for (wireless)
  microphone networks using message passing}.
\newblock In {\it \bibinfo{booktitle}{Acoustic Signal Enhancement; Proceedings
  of IWAENC 2012; International Workshop on}\/} (pp. \bibinfo{pages}{1--4}).
\newblock \bibinfo{organization}{VDE}.
\bibitem[{Heymann et~al.(2016)Heymann, Drude \&
  Haeb-Umbach}]{heymann2016neural}
\bibinfo{author}{Heymann, J.}, \bibinfo{author}{Drude, L.}, \&
  \bibinfo{author}{Haeb-Umbach, R.} (\bibinfo{year}{2016}).
\newblock \bibinfo{title}{Neural network based spectral mask estimation for
  acoustic beamforming}.
\newblock In {\it \bibinfo{booktitle}{Acoustics, Speech and Signal Processing
  (ICASSP), 2016 IEEE International Conference on}\/} (pp.
  \bibinfo{pages}{196--200}).
\newblock \bibinfo{organization}{IEEE}.
\bibitem[{Higuchi et~al.(2016)Higuchi, Ito, Yoshioka \&
  Nakatani}]{higuchi2016robust}
\bibinfo{author}{Higuchi, T.}, \bibinfo{author}{Ito, N.},
  \bibinfo{author}{Yoshioka, T.}, \& \bibinfo{author}{Nakatani, T.}
  (\bibinfo{year}{2016}).
\newblock \bibinfo{title}{Robust {MVDR} beamforming using time-frequency masks
  for online/offline {ASR} in noise}.
\newblock In {\it \bibinfo{booktitle}{Acoustics, Speech and Signal Processing
  (ICASSP), 2016 IEEE International Conference on}\/} (pp.
  \bibinfo{pages}{5210--5214}).
\newblock \bibinfo{organization}{IEEE}.
\bibitem[{Higuchi et~al.(2018)Higuchi, Kinoshita, Ito, Karita \&
  Nakatani}]{higuchi2018frame}
\bibinfo{author}{Higuchi, T.}, \bibinfo{author}{Kinoshita, K.},
  \bibinfo{author}{Ito, N.}, \bibinfo{author}{Karita, S.}, \&
  \bibinfo{author}{Nakatani, T.} (\bibinfo{year}{2018}).
\newblock \bibinfo{title}{Frame-by-frame closed-form update for mask-based
  adaptive {MVDR} beamforming}.
\newblock In {\it \bibinfo{booktitle}{2018 IEEE International Conference on
  Acoustics, Speech and Signal Processing (ICASSP)}\/} (pp.
  \bibinfo{pages}{531--535}).
\newblock \bibinfo{organization}{IEEE}.
\bibitem[{Huang et~al.(2015)Huang, Kim, Hasegawa-Johnson \&
  Smaragdis}]{huang2015joint}
\bibinfo{author}{Huang, P.-S.}, \bibinfo{author}{Kim, M.},
  \bibinfo{author}{Hasegawa-Johnson, M.}, \& \bibinfo{author}{Smaragdis, P.}
  (\bibinfo{year}{2015}).
\newblock \bibinfo{title}{Joint optimization of masks and deep recurrent neural
  networks for monaural source separation}.
\newblock {\it \bibinfo{journal}{IEEE/ACM Trans. Audio, Speech, Lang.
  Process.}\/},  {\it \bibinfo{volume}{23}\/}, \bibinfo{pages}{2136--2147}.
\bibitem[{Jayaprakasam et~al.(2017)Jayaprakasam, Rahim \&
  Leow}]{jayaprakasam2017distributed}
\bibinfo{author}{Jayaprakasam, S.}, \bibinfo{author}{Rahim, S. K.~A.}, \&
  \bibinfo{author}{Leow, C.~Y.} (\bibinfo{year}{2017}).
\newblock \bibinfo{title}{Distributed and collaborative beamforming in wireless
  sensor networks: Classifications, trends, and research directions}.
\newblock {\it \bibinfo{journal}{IEEE Communications Surveys \& Tutorials}\/},
  {\it \bibinfo{volume}{19}\/}, \bibinfo{pages}{2092--2116}.
\bibitem[{Jiang et~al.(2014)Jiang, Wang, Liu \& Feng}]{jiang2014binaural}
\bibinfo{author}{Jiang, Y.}, \bibinfo{author}{Wang, D.}, \bibinfo{author}{Liu,
  R.}, \& \bibinfo{author}{Feng, Z.} (\bibinfo{year}{2014}).
\newblock \bibinfo{title}{Binaural classification for reverberant speech
  segregation using deep neural networks}.
\newblock {\it \bibinfo{journal}{IEEE/ACM Transactions on Audio, Speech and
  Language Processing (TASLP)}\/},  {\it \bibinfo{volume}{22}\/},
  \bibinfo{pages}{2112--2121}.
\bibitem[{Knapp \& Carter(1976)}]{knapp1976generalized}
\bibinfo{author}{Knapp, C.}, \& \bibinfo{author}{Carter, G.}
  (\bibinfo{year}{1976}).
\newblock \bibinfo{title}{The generalized correlation method for estimation of
  time delay}.
\newblock {\it \bibinfo{journal}{IEEE transactions on acoustics, speech, and
  signal processing}\/},  {\it \bibinfo{volume}{24}\/},
  \bibinfo{pages}{320--327}.
\bibitem[{Koutrouvelis et~al.(2018)Koutrouvelis, Sherson, Heusdens \&
  Hendriks}]{koutrouvelis2018low}
\bibinfo{author}{Koutrouvelis, A.~I.}, \bibinfo{author}{Sherson, T.~W.},
  \bibinfo{author}{Heusdens, R.}, \& \bibinfo{author}{Hendriks, R.~C.}
  (\bibinfo{year}{2018}).
\newblock \bibinfo{title}{A low-cost robust distributed linearly constrained
  beamformer for wireless acoustic sensor networks with arbitrary topology}.
\newblock {\it \bibinfo{journal}{IEEE/ACM Transactions on Audio, Speech and
  Language Processing (TASLP)}\/},  {\it \bibinfo{volume}{26}\/},
  \bibinfo{pages}{1434--1448}.
\bibitem[{Lu et~al.(2013)Lu, Tsao, Matsuda \& Hori}]{lu2013speech}
\bibinfo{author}{Lu, X.}, \bibinfo{author}{Tsao, Y.}, \bibinfo{author}{Matsuda,
  S.}, \& \bibinfo{author}{Hori, C.} (\bibinfo{year}{2013}).
\newblock \bibinfo{title}{Speech enhancement based on deep denoising
  autoencoder.}
\newblock In {\it \bibinfo{booktitle}{Interspeech}\/} (pp.
  \bibinfo{pages}{436--440}).
\bibitem[{Luo \& Mesgarani(2019)}]{luo2019conv}
\bibinfo{author}{Luo, Y.}, \& \bibinfo{author}{Mesgarani, N.}
  (\bibinfo{year}{2019}).
\newblock \bibinfo{title}{Conv-tasnet: Surpassing ideal time--frequency
  magnitude masking for speech separation}.
\newblock {\it \bibinfo{journal}{IEEE/ACM transactions on audio, speech, and
  language processing}\/},  {\it \bibinfo{volume}{27}\/},
  \bibinfo{pages}{1256--1266}.
\bibitem[{Mallat(1989)}]{mallat1989theory}
\bibinfo{author}{Mallat, S.~G.} (\bibinfo{year}{1989}).
\newblock \bibinfo{title}{A theory for multiresolution signal decomposition:
  the wavelet representation}.
\newblock {\it \bibinfo{journal}{IEEE transactions on pattern analysis and
  machine intelligence}\/},  {\it \bibinfo{volume}{11}\/},
  \bibinfo{pages}{674--693}.
\bibitem[{Markovich-Golan et~al.(2012)Markovich-Golan, Gannot \&
  Cohen}]{markovich2012distributed}
\bibinfo{author}{Markovich-Golan, S.}, \bibinfo{author}{Gannot, S.}, \&
  \bibinfo{author}{Cohen, I.} (\bibinfo{year}{2012}).
\newblock \bibinfo{title}{Distributed multiple constraints generalized sidelobe
  canceler for fully connected wireless acoustic sensor networks}.
\newblock {\it \bibinfo{journal}{IEEE Transactions on Audio, Speech, and
  Language Processing}\/},  {\it \bibinfo{volume}{21}\/},
  \bibinfo{pages}{343--356}.
\bibitem[{Nakatani et~al.(2017)Nakatani, Ito, Higuchi, Araki \&
  Kinoshita}]{nakatani2017integrating}
\bibinfo{author}{Nakatani, T.}, \bibinfo{author}{Ito, N.},
  \bibinfo{author}{Higuchi, T.}, \bibinfo{author}{Araki, S.}, \&
  \bibinfo{author}{Kinoshita, K.} (\bibinfo{year}{2017}).
\newblock \bibinfo{title}{Integrating {DNN}-based and spatial clustering-based
  mask estimation for robust {MVDR} beamforming}.
\newblock In {\it \bibinfo{booktitle}{Acoustics, Speech and Signal Processing
  (ICASSP), 2017 IEEE International Conference on}\/} (pp.
  \bibinfo{pages}{286--290}).
\newblock \bibinfo{organization}{IEEE}.
\bibitem[{Ng et~al.(2001)Ng, Jordan \& Weiss}]{ng2001spectral}
\bibinfo{author}{Ng, A.~Y.}, \bibinfo{author}{Jordan, M.~I.}, \&
  \bibinfo{author}{Weiss, Y.} (\bibinfo{year}{2001}).
\newblock \bibinfo{title}{{On spectral clustering: Analysis and an algorithm}}.
\newblock In {\it \bibinfo{booktitle}{NIPS}\/}.
\bibitem[{O'Connor \& Kleijn(2014)}]{o2014diffusion}
\bibinfo{author}{O'Connor, M.}, \& \bibinfo{author}{Kleijn, W.~B.}
  (\bibinfo{year}{2014}).
\newblock \bibinfo{title}{Diffusion-based distributed {MVDR} beamformer}.
\newblock In {\it \bibinfo{booktitle}{Acoustics, Speech and Signal Processing
  (ICASSP), 2014 IEEE International Conference on}\/} (pp.
  \bibinfo{pages}{810--814}).
\newblock \bibinfo{organization}{IEEE}.
\bibitem[{O'Connor et~al.(2016)O'Connor, Kleijn \&
  Abhayapala}]{o2016distributed}
\bibinfo{author}{O'Connor, M.}, \bibinfo{author}{Kleijn, W.~B.}, \&
  \bibinfo{author}{Abhayapala, T.} (\bibinfo{year}{2016}).
\newblock \bibinfo{title}{Distributed sparse {MVDR} beamforming using the
  bi-alternating direction method of multipliers}.
\newblock In {\it \bibinfo{booktitle}{Acoustics, Speech and Signal Processing
  (ICASSP), 2016 IEEE International Conference on}\/} (pp.
  \bibinfo{pages}{106--110}).
\newblock \bibinfo{organization}{IEEE}.
\bibitem[{Pandey \& Wang(2019)}]{pandey2019new}
\bibinfo{author}{Pandey, A.}, \& \bibinfo{author}{Wang, D.}
  (\bibinfo{year}{2019}).
\newblock \bibinfo{title}{A new framework for cnn-based speech enhancement in
  the time domain}.
\newblock {\it \bibinfo{journal}{IEEE/ACM Transactions on Audio, Speech, and
  Language Processing}\/},  {\it \bibinfo{volume}{27}\/},
  \bibinfo{pages}{1179--1188}.
\bibitem[{Pariente et~al.(2020)Pariente, Cornell, Deleforge \&
  Vincent}]{pariente2020filterbank}
\bibinfo{author}{Pariente, M.}, \bibinfo{author}{Cornell, S.},
  \bibinfo{author}{Deleforge, A.}, \& \bibinfo{author}{Vincent, E.}
  (\bibinfo{year}{2020}).
\newblock \bibinfo{title}{Filterbank design for end-to-end speech separation}.
\newblock In {\it \bibinfo{booktitle}{ICASSP 2020-2020 IEEE International
  Conference on Acoustics, Speech and Signal Processing (ICASSP)}\/} (pp.
  \bibinfo{pages}{6364--6368}).
\newblock \bibinfo{organization}{IEEE}.
\bibitem[{Qi et~al.(2019)Qi, Du, Siniscalchi \& Lee}]{qi2019theory}
\bibinfo{author}{Qi, J.}, \bibinfo{author}{Du, J.},
  \bibinfo{author}{Siniscalchi, S.~M.}, \& \bibinfo{author}{Lee, C.-H.}
  (\bibinfo{year}{2019}).
\newblock \bibinfo{title}{A theory on deep neural network based
  vector-to-vector regression with an illustration of its expressive power in
  speech enhancement}.
\newblock {\it \bibinfo{journal}{IEEE/ACM Transactions on Audio, Speech, and
  Language Processing}\/},  {\it \bibinfo{volume}{27}\/},
  \bibinfo{pages}{1932--1943}.
\bibitem[{Rix et~al.(2001)Rix, Beerends, Hollier \&
  Hekstra}]{rix2001perceptual}
\bibinfo{author}{Rix, A.~W.}, \bibinfo{author}{Beerends, J.~G.},
  \bibinfo{author}{Hollier, M.~P.}, \& \bibinfo{author}{Hekstra, A.~P.}
  (\bibinfo{year}{2001}).
\newblock \bibinfo{title}{Perceptual evaluation of speech quality {(PESQ)}-a
  new method for speech quality assessment of telephone networks and codecs}.
\newblock In {\it \bibinfo{booktitle}{Proc. IEEE Int. Conf. Acoust., Speech,
  Signal Process.}\/} (pp. \bibinfo{pages}{749--752}).
\bibitem[{Sep{\'u}lveda et~al.(2013)Sep{\'u}lveda, Guido \&
  Castellanos-Dominguez}]{sepulveda2013estimation}
\bibinfo{author}{Sep{\'u}lveda, A.}, \bibinfo{author}{Guido, R.~C.}, \&
  \bibinfo{author}{Castellanos-Dominguez, G.} (\bibinfo{year}{2013}).
\newblock \bibinfo{title}{Estimation of relevant time--frequency features using
  kendall coefficient for articulator position inference}.
\newblock {\it \bibinfo{journal}{Speech communication}\/},  {\it
  \bibinfo{volume}{55}\/}, \bibinfo{pages}{99--110}.
\bibitem[{Taal et~al.(2011)Taal, Hendriks, Heusdens \&
  Jensen}]{taal2011algorithm}
\bibinfo{author}{Taal, C.~H.}, \bibinfo{author}{Hendriks, R.~C.},
  \bibinfo{author}{Heusdens, R.}, \& \bibinfo{author}{Jensen, J.}
  (\bibinfo{year}{2011}).
\newblock \bibinfo{title}{An algorithm for intelligibility prediction of
  time--frequency weighted noisy speech}.
\newblock {\it \bibinfo{journal}{IEEE Trans. Audio, Speech, Lang. Process.}\/},
   {\it \bibinfo{volume}{19}\/}, \bibinfo{pages}{2125--2136}.
\bibitem[{Taherian et~al.(2019)Taherian, Wang \& Wang}]{taherian2019deep}
\bibinfo{author}{Taherian, H.}, \bibinfo{author}{Wang, Z.-Q.}, \&
  \bibinfo{author}{Wang, D.} (\bibinfo{year}{2019}).
\newblock \bibinfo{title}{Deep learning based multi-channel speaker recognition
  in noisy and reverberant environments}.
\newblock {\it \bibinfo{journal}{Proc. Interspeech 2019}\/},  (pp.
  \bibinfo{pages}{4070--4074}).
\bibitem[{Tan et~al.(2018)Tan, Chen \& Wang}]{tan2018gated}
\bibinfo{author}{Tan, K.}, \bibinfo{author}{Chen, J.}, \&
  \bibinfo{author}{Wang, D.} (\bibinfo{year}{2018}).
\newblock \bibinfo{title}{Gated residual networks with dilated convolutions for
  monaural speech enhancement}.
\newblock {\it \bibinfo{journal}{IEEE/ACM transactions on audio, speech, and
  language processing}\/},  {\it \bibinfo{volume}{27}\/},
  \bibinfo{pages}{189--198}.
\bibitem[{Tan \& Wang(2019)}]{tan2019learning}
\bibinfo{author}{Tan, K.}, \& \bibinfo{author}{Wang, D.}
  (\bibinfo{year}{2019}).
\newblock \bibinfo{title}{Learning complex spectral mapping with gated
  convolutional recurrent networks for monaural speech enhancement}.
\newblock {\it \bibinfo{journal}{IEEE/ACM Transactions on Audio, Speech, and
  Language Processing}\/},  {\it \bibinfo{volume}{28}\/},
  \bibinfo{pages}{380--390}.
\bibitem[{Tavakoli et~al.(2016)Tavakoli, Jensen, Christensen \&
  Benesty}]{tavakoli2016framework}
\bibinfo{author}{Tavakoli, V.~M.}, \bibinfo{author}{Jensen, J.~R.},
  \bibinfo{author}{Christensen, M.~G.}, \& \bibinfo{author}{Benesty, J.}
  (\bibinfo{year}{2016}).
\newblock \bibinfo{title}{A framework for speech enhancement with ad hoc
  microphone arrays}.
\newblock {\it \bibinfo{journal}{IEEE/ACM Transactions on Audio, Speech and
  Language Processing (TASLP)}\/},  {\it \bibinfo{volume}{24}\/},
  \bibinfo{pages}{1038--1051}.
\bibitem[{Tavakoli et~al.(2017)Tavakoli, Jensen, Heusdens, Benesty \&
  Christensen}]{tavakoli2017distributed}
\bibinfo{author}{Tavakoli, V.~M.}, \bibinfo{author}{Jensen, J.~R.},
  \bibinfo{author}{Heusdens, R.}, \bibinfo{author}{Benesty, J.}, \&
  \bibinfo{author}{Christensen, M.~G.} (\bibinfo{year}{2017}).
\newblock \bibinfo{title}{Distributed max-{SINR} speech enhancement with ad hoc
  microphone arrays}.
\newblock In {\it \bibinfo{booktitle}{Acoustics, Speech and Signal Processing
  (ICASSP), 2017 IEEE International Conference on}\/} (pp.
  \bibinfo{pages}{151--155}).
\newblock \bibinfo{organization}{IEEE}.
\bibitem[{Tu et~al.(2017)Tu, Du, Sun \& Lee}]{tu2017lstm}
\bibinfo{author}{Tu, Y.-H.}, \bibinfo{author}{Du, J.}, \bibinfo{author}{Sun,
  L.}, \& \bibinfo{author}{Lee, C.-H.} (\bibinfo{year}{2017}).
\newblock \bibinfo{title}{{LSTM}-based iterative mask estimation and
  post-processing for multi-channel speech enhancement}.
\newblock In {\it \bibinfo{booktitle}{Asia-Pacific Signal and Information
  Processing Association Annual Summit and Conference (APSIPA ASC), 2017}\/}
  (pp. \bibinfo{pages}{488--491}).
\newblock \bibinfo{organization}{IEEE}.
\bibitem[{Vincent et~al.(2006)Vincent, Gribonval \&
  F{\'e}votte}]{vincent2006performance}
\bibinfo{author}{Vincent, E.}, \bibinfo{author}{Gribonval, R.}, \&
  \bibinfo{author}{F{\'e}votte, C.} (\bibinfo{year}{2006}).
\newblock \bibinfo{title}{Performance measurement in blind audio source
  separation}.
\newblock {\it \bibinfo{journal}{IEEE Trans. Audio, Speech, Lang. Process.}\/},
   {\it \bibinfo{volume}{14}\/}, \bibinfo{pages}{1462--1469}.
\bibitem[{Wang \& Chen(2018)}]{wang2018supervised}
\bibinfo{author}{Wang, D.}, \& \bibinfo{author}{Chen, J.}
  (\bibinfo{year}{2018}).
\newblock \bibinfo{title}{Supervised speech separation based on deep learning:
  An overview}.
\newblock {\it \bibinfo{journal}{IEEE/ACM Transactions on Audio, Speech, and
  Language Processing}\/}, .
\bibitem[{Wang \& Cavallaro(2018)}]{wang2018pseudo}
\bibinfo{author}{Wang, L.}, \& \bibinfo{author}{Cavallaro, A.}
  (\bibinfo{year}{2018}).
\newblock \bibinfo{title}{Pseudo-determined blind source separation for ad-hoc
  microphone networks}.
\newblock {\it \bibinfo{journal}{IEEE/ACM Transactions on Audio, Speech, and
  Language Processing}\/},  {\it \bibinfo{volume}{26}\/},
  \bibinfo{pages}{981--994}.
\bibitem[{Wang et~al.(2015)Wang, Hon, Reiss \& Cavallaro}]{wang2015self}
\bibinfo{author}{Wang, L.}, \bibinfo{author}{Hon, T.-K.},
  \bibinfo{author}{Reiss, J.~D.}, \& \bibinfo{author}{Cavallaro, A.}
  (\bibinfo{year}{2015}).
\newblock \bibinfo{title}{Self-localization of ad-hoc arrays using time
  difference of arrivals}.
\newblock {\it \bibinfo{journal}{IEEE Transactions on Signal Processing}\/},
  {\it \bibinfo{volume}{64}\/}, \bibinfo{pages}{1018--1033}.
\bibitem[{Wang et~al.(2019)Wang, Tan et~al.}]{wang2019bridging}
\bibinfo{author}{Wang, P.}, \bibinfo{author}{Tan, K.} et~al.
  (\bibinfo{year}{2019}).
\newblock \bibinfo{title}{Bridging the gap between monaural speech enhancement
  and recognition with distortion-independent acoustic modeling}.
\newblock {\it \bibinfo{journal}{IEEE/ACM Transactions on Audio, Speech, and
  Language Processing}\/},  {\it \bibinfo{volume}{28}\/},
  \bibinfo{pages}{39--48}.
\bibitem[{Wang et~al.(2014)Wang, Narayanan \& Wang}]{wang2014training}
\bibinfo{author}{Wang, Y.}, \bibinfo{author}{Narayanan, A.}, \&
  \bibinfo{author}{Wang, D.~L.} (\bibinfo{year}{2014}).
\newblock \bibinfo{title}{On training targets for supervised speech
  separation}.
\newblock {\it \bibinfo{journal}{IEEE/ACM Trans. Audio, Speech, Lang.
  Process.}\/},  {\it \bibinfo{volume}{22}\/}, \bibinfo{pages}{1849--1858}.
\bibitem[{Wang \& Wang(2013)}]{wang2013towards}
\bibinfo{author}{Wang, Y.}, \& \bibinfo{author}{Wang, D.~L.}
  (\bibinfo{year}{2013}).
\newblock \bibinfo{title}{Towards scaling up classification-based speech
  separation}.
\newblock {\it \bibinfo{journal}{IEEE Trans. Audio, Speech, Lang. Process.}\/},
   {\it \bibinfo{volume}{21}\/}, \bibinfo{pages}{1381--1390}.
\bibitem[{Wang \& Wang(2018)}]{wang2018all}
\bibinfo{author}{Wang, Z.-Q.}, \& \bibinfo{author}{Wang, D.}
  (\bibinfo{year}{2018}).
\newblock \bibinfo{title}{All-neural multichannel speech enhancement}.
\newblock {\it \bibinfo{journal}{to appear in Interspeech}\/}, .
\bibitem[{Wang et~al.(2018)Wang, Zhang \& Wang}]{wang2018robust}
\bibinfo{author}{Wang, Z.-Q.}, \bibinfo{author}{Zhang, X.}, \&
  \bibinfo{author}{Wang, D.} (\bibinfo{year}{2018}).
\newblock \bibinfo{title}{Robust speaker localization guided by deep
  learning-based time-frequency masking}.
\newblock {\it \bibinfo{journal}{IEEE/ACM Transactions on Audio, Speech, and
  Language Processing}\/},  {\it \bibinfo{volume}{27}\/},
  \bibinfo{pages}{178--188}.
\bibitem[{Weninger et~al.(2015)Weninger, Erdogan, Watanabe, Vincent, Le~Roux,
  Hershey \& Schuller}]{weninger2015speech}
\bibinfo{author}{Weninger, F.}, \bibinfo{author}{Erdogan, H.},
  \bibinfo{author}{Watanabe, S.}, \bibinfo{author}{Vincent, E.},
  \bibinfo{author}{Le~Roux, J.}, \bibinfo{author}{Hershey, J.~R.}, \&
  \bibinfo{author}{Schuller, B.} (\bibinfo{year}{2015}).
\newblock \bibinfo{title}{Speech enhancement with lstm recurrent neural
  networks and its application to noise-robust asr}.
\newblock In {\it \bibinfo{booktitle}{International Conference on Latent
  Variable Analysis and Signal Separation}\/} (pp. \bibinfo{pages}{91--99}).
\newblock \bibinfo{organization}{Springer}.
\bibitem[{Williamson et~al.(2016)Williamson, Wang \&
  Wang}]{williamson2016complex}
\bibinfo{author}{Williamson, D.~S.}, \bibinfo{author}{Wang, Y.}, \&
  \bibinfo{author}{Wang, D.~L.} (\bibinfo{year}{2016}).
\newblock \bibinfo{title}{Complex ratio masking for monaural speech
  separation}.
\newblock {\it \bibinfo{journal}{IEEE/ACM Trans. Audio, Speech, Lang.
  Process.}\/},  {\it \bibinfo{volume}{24}\/}, \bibinfo{pages}{483--492}.
\bibitem[{Xiao et~al.(2017)Xiao, Zhao, Jones, Chng \& Li}]{xiao2017time}
\bibinfo{author}{Xiao, X.}, \bibinfo{author}{Zhao, S.}, \bibinfo{author}{Jones,
  D.~L.}, \bibinfo{author}{Chng, E.~S.}, \& \bibinfo{author}{Li, H.}
  (\bibinfo{year}{2017}).
\newblock \bibinfo{title}{On time-frequency mask estimation for {MVDR}
  beamforming with application in robust speech recognition}.
\newblock In {\it \bibinfo{booktitle}{Acoustics, Speech and Signal Processing
  (ICASSP), 2017 IEEE International Conference on}\/} (pp.
  \bibinfo{pages}{3246--3250}).
\newblock \bibinfo{organization}{IEEE}.
\bibitem[{Xu et~al.(2015)Xu, Du, Dai \& Lee}]{xu2015regression}
\bibinfo{author}{Xu, Y.}, \bibinfo{author}{Du, J.}, \bibinfo{author}{Dai,
  L.-R.}, \& \bibinfo{author}{Lee, C.-H.} (\bibinfo{year}{2015}).
\newblock \bibinfo{title}{A regression approach to speech enhancement based on
  deep neural networks}.
\newblock {\it \bibinfo{journal}{IEEE/ACM Trans. Audio, Speech, Lang.
  Process.}\/},  {\it \bibinfo{volume}{23}\/}, \bibinfo{pages}{7--19}.
\bibitem[{Yang \& Zhang(2019)}]{yang2019boosting}
\bibinfo{author}{Yang, Z.}, \& \bibinfo{author}{Zhang, X.-L.}
  (\bibinfo{year}{2019}).
\newblock \bibinfo{title}{Boosting spatial information for deep learning based
  multichannel speaker-independent speech separation in reverberant
  environments}.
\newblock In {\it \bibinfo{booktitle}{2019 Asia-Pacific Signal and Information
  Processing Association Annual Summit and Conference (APSIPA ASC)}\/} (pp.
  \bibinfo{pages}{1506--1510}).
\newblock \bibinfo{organization}{IEEE}.
\bibitem[{Zeng \& Hendriks(2014)}]{zeng2014distributed}
\bibinfo{author}{Zeng, Y.}, \& \bibinfo{author}{Hendriks, R.~C.}
  (\bibinfo{year}{2014}).
\newblock \bibinfo{title}{Distributed delay and sum beamformer for speech
  enhancement via randomized gossip}.
\newblock {\it \bibinfo{journal}{IEEE/ACM Transactions on Audio, Speech, and
  Language Processing}\/},  {\it \bibinfo{volume}{22}\/},
  \bibinfo{pages}{260--273}.
\bibitem[{Zhang et~al.(2018)Zhang, Chepuri, Hendriks \&
  Heusdens}]{zhang2018microphone}
\bibinfo{author}{Zhang, J.}, \bibinfo{author}{Chepuri, S.~P.},
  \bibinfo{author}{Hendriks, R.~C.}, \& \bibinfo{author}{Heusdens, R.}
  (\bibinfo{year}{2018}).
\newblock \bibinfo{title}{Microphone subset selection for {MVDR} beamformer
  based noise reduction}.
\newblock {\it \bibinfo{journal}{IEEE/ACM Transactions on Audio, Speech, and
  Language Processing}\/},  {\it \bibinfo{volume}{26}\/},
  \bibinfo{pages}{550--563}.
\bibitem[{Zhang et~al.(2017)Zhang, Wang \& Wang}]{zhang2017speech}
\bibinfo{author}{Zhang, X.}, \bibinfo{author}{Wang, Z.-Q.}, \&
  \bibinfo{author}{Wang, D.} (\bibinfo{year}{2017}).
\newblock \bibinfo{title}{A speech enhancement algorithm by iterating
  single-and multi-microphone processing and its application to robust {ASR}}.
\newblock In {\it \bibinfo{booktitle}{Acoustics, Speech and Signal Processing
  (ICASSP), 2017 IEEE International Conference on}\/} (pp.
  \bibinfo{pages}{276--280}).
\newblock \bibinfo{organization}{IEEE}.
\bibitem[{Zhang(2018)}]{zhang2018deep}
\bibinfo{author}{Zhang, X.-L.} (\bibinfo{year}{2018}).
\newblock \bibinfo{title}{Deep ad-hoc beamforming}.
\newblock {\it \bibinfo{journal}{arXiv preprint arXiv:1811.01233}\/}, .
  \URLprefix \url{https://arxiv.org/abs/1811.01233v2}.
\bibitem[{Zhang \& Wang(2016{\natexlab{a}})}]{zhang2016boosting}
\bibinfo{author}{Zhang, X.-L.}, \& \bibinfo{author}{Wang, D.}
  (\bibinfo{year}{2016}{\natexlab{a}}).
\newblock \bibinfo{title}{{Boosting contextual information for deep neural
  network based voice activity detection}}.
\newblock {\it \bibinfo{journal}{IEEE/ACM Transactions on Audio, Speech, and
  Language Processing}\/},  {\it \bibinfo{volume}{24}\/},
  \bibinfo{pages}{252--264}.
\bibitem[{Zhang \& Wang(2016{\natexlab{b}})}]{zhang2016deep}
\bibinfo{author}{Zhang, X.-L.}, \& \bibinfo{author}{Wang, D.}
  (\bibinfo{year}{2016}{\natexlab{b}}).
\newblock \bibinfo{title}{A deep ensemble learning method for monaural speech
  separation}.
\newblock {\it \bibinfo{journal}{IEEE/ACM transactions on audio, speech, and
  language processing}\/},  {\it \bibinfo{volume}{24}\/},
  \bibinfo{pages}{967--977}.
\bibitem[{Zhang \& Wu(2013{\natexlab{a}})}]{zhang2013deep}
\bibinfo{author}{Zhang, X.-L.}, \& \bibinfo{author}{Wu, J.}
  (\bibinfo{year}{2013}{\natexlab{a}}).
\newblock \bibinfo{title}{Deep belief networks based voice activity detection}.
\newblock {\it \bibinfo{journal}{IEEE Trans. Audio, Speech, Lang. Process.}\/},
   {\it \bibinfo{volume}{21}\/}, \bibinfo{pages}{697--710}.
\bibitem[{Zhang \& Wu(2013{\natexlab{b}})}]{zhang2013denoising}
\bibinfo{author}{Zhang, X.-L.}, \& \bibinfo{author}{Wu, J.}
  (\bibinfo{year}{2013}{\natexlab{b}}).
\newblock \bibinfo{title}{Denoising deep neural networks based voice activity
  detection}.
\newblock In {\it \bibinfo{booktitle}{the 38th IEEE International Conference on
  Acoustic, Speech, and Signal Processing}\/} (pp. \bibinfo{pages}{853--857}).
\bibitem[{Zheng \& Zhang(2018)}]{zheng2018phase}
\bibinfo{author}{Zheng, N.}, \& \bibinfo{author}{Zhang, X.-L.}
  (\bibinfo{year}{2018}).
\newblock \bibinfo{title}{Phase-aware speech enhancement based on deep neural
  networks}.
\newblock {\it \bibinfo{journal}{IEEE/ACM Transactions on Audio, Speech, and
  Language Processing}\/},  {\it \bibinfo{volume}{27}\/},
  \bibinfo{pages}{63--76}.
\bibitem[{Zheng et~al.(2019)Zheng, Tang \& Zhou}]{zheng2019framework}
\bibinfo{author}{Zheng, X.}, \bibinfo{author}{Tang, Y.~Y.}, \&
  \bibinfo{author}{Zhou, J.} (\bibinfo{year}{2019}).
\newblock \bibinfo{title}{A framework of adaptive multiscale wavelet
  decomposition for signals on undirected graphs}.
\newblock {\it \bibinfo{journal}{IEEE Transactions on Signal Processing}\/},
  {\it \bibinfo{volume}{67}\/}, \bibinfo{pages}{1696--1711}.
\bibitem[{Zhou \& Qian(2018)}]{zhou2018robust}
\bibinfo{author}{Zhou, Y.}, \& \bibinfo{author}{Qian, Y.}
  (\bibinfo{year}{2018}).
\newblock \bibinfo{title}{Robust mask estimation by integrating neural
  network-based and clustering-based approaches for adaptive acoustic
  beamforming}.
\newblock In {\it \bibinfo{booktitle}{Int Conf on Acoustics, Speech, and Signal
  Processing, in press. Google Scholar}\/}.

\end{thebibliography}

\clearpage
\end{document}